\documentstyle[aps,epsfig]{revtex}
\def \ut#1{\rlap{\lower1ex\hbox{$\sim$}}#1{}}
\newcommand{\f}{\begin{equation}}
\newcommand{\ff}{\end{equation}}

\newcommand{\pic}[5]{\raisebox{#3pt}{
\hspace{#4pt}\psfig{file=#1.ps,height=#2pt,silent=}\hspace{#5pt}}}

\newcommand{\kd}[1]{\mathchoice{
\pic{#1}{75}{-35}{0}{0}}{
\pic{#1}{12}{-2}{-3}{2}}{
\pic{#1}{9}{-2}{-3}{1}}{
\pic{#1}{7}{-1}{-1}{0}}}
\begin{document}
\draft
\title{Graphical Evolution of Spin Network States} 
\author{Roumen Borissov\thanks{Present address: Center for Gravitational
Physics and Geometry, The Pennsylvania State University, University Park, PA
16802; e-mail address:borissov@phys.psu.edu}}
\address{Physics Department , Temple University\\ Philadelphia, PA 19122}
\date{\today}
\maketitle

\begin{abstract}
The evolution of spin network states in loop quantum gravity can be 
defined with respect to a time variable, given by the surfaces of 
constant value of an auxiliary scalar field. We regulate the Hamiltonian, 
generating such an evolution, and evaluate its action both on edges and 
on vertices of the spin network states. The analytical computations are 
carried out completely to yield a finite, diffeomorphism invariant result. 
We use techniques from the recoupling theory of colored graphs with 
trivalent vertices to evaluate the graphical part of the Hamiltonian 
action. We show that the action on edges is equivalent to a 
diffeomorphism transformation, while the action on vertices adds new edges 
and re-routes the loops through the vertices. A remaining usresolved 
problem is to take the square root of the infinite-dimmensional matrix 
of the Hamiltonian constraint and to obtain the eigenspectrum of the
``clock field'' Hamiltonian.
\end{abstract}


\section{Introduction}
\label{sec:intro}

In canonical quantum gravity the notion of evolution requires careful
definition because the translations in time direction can be interpreted
as diffeomorphism transformations \cite{MTW}. To be able to talk about 
evolution we can use ``relational constructions". Some physical fields 
can be introduced as a reference frame with respect to which
the evolution can be defined \cite{Kuch}, \cite{Smo}, \cite{RoSmo1}. In loop 
quantum gravity \cite{biblio} Rovelli and Smolin \cite{RoSmo1}, 
\cite{Ro1} define a time variable by the surfaces of constant value of 
an auxiliary scalar field. By fixing a gauge in this construction 
the infinite number of Hamiltonian constraints (one per space point) reduces 
to one constraint which can be interpreted as a Schr\"{o}dinger equation 
and a Hamiltonian operator can be identified. We use the Hamiltonian
obtained in this model to investigate the evolution of quantum
gravitational states. In loop quantum gravity spin network states, 
\cite{RoSmo2},  \cite{ALMMT}, \cite{Baez}, furnish a complete basis 
of quantum kinematical states.

There exists \cite{RoSmo0}, \cite{DePiRo} already a well established 
procedure for expressing different quantities from quantum gravity in 
terms of the loop variables \cite{RoSmo_loop}. Then operator versions of 
the gravitational quantities can be defined by replacing the loop 
variables with the corresponding operators. Thus it is relatively 
straightforward to introduce a loop version for the Hamiltonian operator. 
We use the result from \cite{RoSmo1} and \cite{Ro1} as a starting point 
for our calculations. Our purpose is to determine in detail the way the 
spin network states evolve under the Hamiltonian introduced in 
\cite{RoSmo1}. As we will see the result of the evolution can be split 
into two parts. First there is a multiplicative factor which is finite
and diffeomorphism invariant. Second -- the spin network states evolve 
topologically: The new state is a sum of terms, each term being 
based on the original spin network with an added extra edge of color 1. 
The added edge connects pairs of the original edges, meeting at a vertex. 
Also some change of coloring of the edges occurs such that the new graph 
is a spin network again.

The content of the paper is as follows: In section 2. we define the 
Hamiltonian and regulate it to show that it has well defined action on 
the spin network states. We also introduce a modification in the way the 
loop operators are defined, better suited for our calculations. 
Technically the action of the Hamiltonian operator can be split  into 
analytical and graphical parts. The analytical part includes various 
pre-factors and integrals. The graphical part expresses the topological 
transformations occurring in the spin networks. In section 3. we compute 
the analytical action of the Hamiltonian separately on edges and on 
vertices. We show that to a great extent the action on edges is 
equivalent to diffeomorphism tranformation. Using some techniques from 
recoupling theory of knots and links with trivalent vertices we perform 
the graphical computation of the action of the Hamiltonian in section 4.  
The result from the graphical calculation tells us whether the 
diffeomorphism class of the spin network or the coloring of certain 
edges change. We conclude with discussion of some open issues.

\section{The Hamiltonian of the theory}
\label{sec:ham}

Because of the absence of external time with 
respect to which the evolution can be defined, we need some additional 
construction. As it has been shown in \cite{RoSmo1} to define time we can 
use the physical degrees of freedom of an auxiliary field.
We start by introducing a scalar field $T(x)$. To serve as a clock, this 
field should be monotonically increasing everywhere on the space manifold 
$\Sigma$\footnote{We use the standard notation - $g^{\mu\nu}$ and $q^{ab}$ are
respectively the 4-metric on ${\cal M}=\Sigma\times R$ and the 3-metric on 
$\Sigma$. $a,b,\dots$ are spatial and $i,j,\dots$ - internal indices; they 
all run from 1 to 3.}. 
Then we can use its 3-surfaces of constant value $T(x)={\it const}$ 
to represent the time with respect to which the evolution will be defined. 
The scalar ``clock" field can be incorporated in the theory through the 
standard Klein-Gordon Lagrangian:

\begin{equation}\label{LagT}
{\cal L}_{T}={\mu \over 2}  g^{\mu\nu} (x)\sqrt{-g(x)}{\partial}_{\mu} T(x)
{\partial}_{\nu} T(x).
\end{equation}

In this expression $\mu$ plays the role of a coupling constant between 
the scalar field and the gravitational field. Later on we will treat $T(x)$ 
as time so, from dimensional analysis, the constant $\mu$ should have 
dimensions of energy density. The momentum conjugate to the field $T(x)$ 
will be:

\begin{equation}\label{momentum}
\tilde{\pi}(x)={\partial{\cal L} \over \partial({\partial}_{0}T)}=\mu q 
N{g}^{0\mu}{\partial}_{\mu}T.
\end{equation}

Performing a Legendre transformation we get for the total Lagrangian:

\begin{equation}\label{}
{\cal L}=\tilde{\pi} {\partial}_{0} T - {\ut N}{{\tilde{\pi}}^{2} \over 2\mu} 
- {\ut N}{\mu \over 2}{q}^{2}{q}^{ab}{\partial}_{a}T{\partial}_{b}T - 
\tilde{\pi} {N}^{b}({\partial}_{b}T) + 
{\cal L}_{Gravity}
\end{equation}

The Euler-Lagrange equations we obtain from this Lagrangian are:

\begin{equation}\label{eq1}
{\delta {\cal L}\over \delta\tilde{\pi}}={\partial}_{0} T - {\ut N}
{{\tilde{\pi}} \over \mu} - {N}^{b}({\partial}_{b}T)= 0,
\end{equation}
and 

\begin{equation}\label{eq2}
{\delta {\cal L} \over \delta T}= -{\partial}_{0} \tilde{\pi }+ 
\mu {\partial}_{b}({\ut N} \stackrel{\approx}{q}^{bc}{\partial}_{c}T) + 
{\partial}_{b}(N^{b} \tilde{\pi} ).
\end{equation}

At this point we impose a gauge fixing, restricting the freedom of 
choosing the time coordinate. The gauge we use is ${\partial}_{a}T(x)=0$ 
which, because of Eq.\ (\ref{eq1}), implies that:
\[
{\partial}_{0} T = {\ut N}{{\tilde{\pi}} \over \mu}.
\]
Thus the lapse function $\ut{N}(x)$ should satisfy the relation:

\begin{equation}\label{10-fix}
\ut{N}(x) = { a(t)\mu \over \tilde{\pi}(x)},
\end{equation}
where $a(t)$ is an arbitrary function of (the coordinate) time. For the 
Hamiltonian constraint we get:

\[
\stackrel{\approx}{{\cal C}}(x) = {{\tilde{\pi}}^{2} \over 2\mu} + 
\stackrel{\approx}{{\cal C}}_{G}(x),
\]
where 
\[
\stackrel{\approx}{{\cal C}}_{G}(x)=  \epsilon_{ijk} \tilde{E}^{ai}
\tilde{E}^{bj}F_{ab}^k - \Lambda  q = {\cal C}_{Einstein} - \Lambda q
\]
is the gravitational Hamiltonian constraint in terms of the Ashtekar 
variables and $\Lambda$ is 
the cosmological constant. According to the general prescription for gauge 
fixing in constraint systems we have to compute the Poisson bracket between 
the gauge and the existing constraints to check for secondary constraints. 
Thus we get:
\[
\{\stackrel{\approx}{{\cal C}}(x'),{\partial}_{a}T(x)\} = {\partial}_{a}
\{ {{\tilde{\pi}}^{2}(x') \over 2\mu}, T(x)\} = {\partial}_{a}
[{\tilde{\pi}(x') \over \mu} {\delta}^{3}(x',x)].
\]
The above expression vanishes when we smear the Hamiltonian constraint
$\stackrel{\approx}{{\cal C}}(x')$ using $\ut{N}(x)$ from eq.Eq.\ (\ref{10-fix}).
Thus the only constraint which remains to be imposed on the wave 
functionals of the theory is the integral of the Hamiltonian constraint 
with the lapse 
function:

\[
\int_{\Sigma}{{d}^{3}}x \ut{N}(x)\stackrel{\approx}{{\cal C}}(x) = 
a(t) \mu \int_{\Sigma}{{{d}^{3}}x \over \tilde{\pi}(x)} \left( {\tilde{\pi}(x) 
\over \sqrt{2\mu}} + \sqrt{- \stackrel{\approx}{{\cal C}_{G}}(x)} \right)
\left( {\tilde{\pi}(x) \over \sqrt{2\mu}} -
 \sqrt{- \stackrel{\approx}{{\cal C}_{G}}(x)} \right) \cong 0.
\]

Because we are going to impose this integral as a constraint operator, we can 
think of ${\tilde{\pi}(x)}$ as being equal to 
$ \sqrt{ -2\mu\stackrel{\approx}{{\cal C}_{G}}(x)}$ Thus the expression in the 
first parentheses can be replaced by $\sqrt{{2 \over \mu}}\tilde{\pi}(x)$ 
and the whole integral reduces to:

\begin{equation}\label{const}
a(t)\int_{\Sigma}{{d}^{3}}x\left( \tilde{\pi}(x)- \sqrt{- 2\mu
\stackrel{\approx}{{\cal C}_{G}}(x)} \right) \cong 0.
\end{equation}

Note that we assume that the Hamiltonian constraint ${\cal C}_{G}(x)$ satisfies
the weak energy condition \cite{KuTo}, which in this case requires that 
${\cal C}_{G}(x) \leq 0$. 

In the process of quantization we promote this constraint into an operator 
equation. We require that in the loop representation the spin network
states, depending also on the clock variable $T$, are annihilated
by the constraint operator:

\begin{equation}\label{anih}
\langle S,T| a(t)\int_{\Sigma}{{d}^{3}}x\left(\hat{ \tilde{\pi}}(x)- 
\sqrt{- 2\mu\hat{\stackrel{\approx}{{\cal C}_{G}}}(x)} \right)=0.
\end{equation}

We interpret the integral:

\[
\int_{\Sigma}{{d}^{3}}x\hat{\tilde{\pi}}(x)
\]
as $i\hbar$ times a derivative with respect to the ``clock" field and thus we 
arrive at the Schr\"{o}dinger equation:

\begin{equation}\label{Schro}
i\hbar{\partial \over \partial T} \langle S, T| =  \langle S, T| \hat{H}
\end{equation}
where the Hamiltonian $\hat{H}$ corresponds to the classical observable
 \footnote{The same result can be obtained \cite{Ro1} by imposing the gauge fixing
at lagrangian level and after that performing the Legendre transformation.}(we 
have restored the constants):

\begin{equation}\label{unregH}
H = {\sqrt{ \mu} \over  \sqrt{4 \pi G}}\int_{\Sigma}{d^{3}x}\sqrt{ - 
C(x) + \Lambda q(x)} . \end{equation}
where

\begin{equation}\label{unregC}
- C(x) + \Lambda q(x)= - \epsilon_{ijk}  \left\{F_{ab}^{k} \widetilde{E}^{ai}
\widetilde{E}^{bj}  - {1 \over 3!} \Lambda \epsilon_{abc}  \widetilde{E}^{ai} \widetilde{E}^{bj} \widetilde{E}^{ck}  \right\}.
\end{equation}
 We will omit the overall factor of 
$(\sqrt{\mu} / \sqrt{4\pi G})$ in what follows, as it is not 
important for our considerations. 

Because of the product of distributional quantities in the Hamiltonian 
we have to regulate the above expression. Using an arbitrary, e.g. Euclidian, 
background metric $h_{ab}(x)$ we divide the space manifold $\Sigma$ into 
cubes of size $L$, labeled $R_{I}$.  In the calculations we will let $L$ 
to go to zero as the number of cubes goes to infinity. Then we can write 
Eq.\ (\ref{unregH}) as

\begin{equation}
H = \lim_{L \rightarrow 0}{\sum_{I}^{\infty}}  \int_{R_{I}}{d^{3}x}
\sqrt{h(x)} \sqrt{ {- C(x) + \Lambda q(x) \over h(x)}}
= \lim_{L \rightarrow 0}{\sum_{I}^{\infty}}  L^{3}
\sqrt{ {- C(x_{I}) + \Lambda q(x_{I}) \over h(x_{I})}},
\end{equation}
where $h(x)$ is the determinant of the background metric, and $x_{I}$ in 
the last expression is a point in the cube $R_{I}$. Now putting
$L^{3}$ inside the square root and going back to integral form we get:

\begin{equation}\label{basic_reg}
H = \lim_{L \rightarrow 0}{\sum_{I}^{\infty}} 
\sqrt{ L^{3} \int_{R_{I}}{d^{3}x \over \sqrt{h(x)} } \left( {- C(x) + 
\Lambda q(x) } \right)} = \lim_{L \rightarrow 0}{\sum_{I}^{\infty}} 
\sqrt{ -  C_{I} + \Lambda V^{2}_{I}}.
\end{equation}
All of the above manipulations are correct in the limit $L \rightarrow 
0$. The last step  in Eq.\ (\ref{basic_reg}) is a definition of $C_{I}$.
$V_{I}$ is the classical volume of the $I$-th cube.

Let us now focus on the first term of the expression under the square 
root in equation Eq.\ (\ref{basic_reg}). We introduce a regulating point-splitting 
function $f_{\delta}(\tilde{x},y)$ which is a density of weight one with 
respect to its first argument. This function satisfies the
requirement that for any smooth function $\phi (x)$:

\begin{equation}
\lim_{\delta\rightarrow 0}{\int_{R_{I} \ni y }{d^{3}x \phi (x)
f_{\delta}(\tilde{x},y)}}= \phi (y).
\end{equation}.

Using the regulating function we can write $C_{I}$ as:

\begin{equation}\label{reg1}
C_{I} =  \lim_{\delta\rightarrow 0} \epsilon_{ijk} L^{3} 
\int_{R_{I}}{d^{3}x \over \sqrt{h(x)} }
F_{ab}^{k}(x)\int_{R_{I}} d^{3}y  
f_{\delta}(\tilde{x},y)\widetilde{E}^{ai}(y)
\int_{R_{I}} d^{3}z  f_{\delta}(\tilde{x},z)\widetilde{E}^{bj}(z).
\end{equation}
To ensure proper contractions of the internal indices in the last expression
we connect the points $x$, $y$, and $z$ with holonomies of the Ashtekar 
connection along some smooth paths $\gamma_1$ and $\gamma_2$ connecting 
the points. We 
also use the identity $\epsilon_{ijk} = - 4 {\rm Tr}[ \tau_{i} \tau_{j} 
\tau_{k}]$ to write Eq.\ (\ref{reg1}) as:

\begin{eqnarray}\label{reg2}
\nonumber
C_{I} =- \lim_{\delta\rightarrow 0} 4 L^{3} 
\int_{R_{I}}{d^{3}x \over \sqrt{h(x)} } \int_{R_{I}} d^{3}y  
f_{\delta}(\tilde{x},y) 
\int_{R_{I}} d^{3}z  f_{\delta}(\tilde{x},z) 
\times \\
\times  {\rm Tr}[F_{ab}^{k}(x) \tau_{k}U_{\gamma_1}(x,y)\widetilde{E}^{ai}(y) 
\tau_{i} U_{\gamma_{1}^{-1}}(y,x) U_{\gamma_2}(x,z)
\widetilde{E}^{bj}(z) \tau_{j}U_{\gamma_{2}^{-1}}(z,x)].
\end{eqnarray}
To complete the regularization we also replace the curvature 
$F_{ab}^{k}(x)$ by its approximation by a holonomy:

\[
F_{ab}^{k}(x) \tau_{k} =  \lim_{\epsilon \rightarrow 0} {1 \over 2 
\epsilon^{2}} U(\gamma^{\epsilon^{2}}_{x,[\hat{a}\hat{b}]}), 
\]
where $\gamma^{\epsilon^{2}}_{x,[\hat{a}\hat{b}]}$ is a loop with area 
$\epsilon^{2}$ in the $(\hat{a},\hat{b})$ coordinate plane, based at the
point  $x$. We have included explicitly the antisymmetrization with 
respect to $\hat{a}$ and $\hat{b}$ to ensure the vanishing of first term 
in the expansion of the holonomy   
$U(\gamma^{\epsilon^{2}}_{x,[\hat{a}\hat{b}]})$ in powers of 
$\epsilon^{2}$. Thus for $C_{I}$ we get:

\begin{eqnarray*}
C_{I} = - \lim_{\epsilon \rightarrow 0} \lim_{\delta\rightarrow 0} 
{L^{3} \over 2 \epsilon^{2}} 
\int_{R_{I}}{d^{3}x \over \sqrt{h(x)} } \int_{R_{I}} d^{3}y  
f_{\delta}(\tilde{x},y) \int_{R_{I}} d^{3}z f_{\delta}(\tilde{x},z)  
\times \\ \times  
{\rm Tr}[U(\gamma^{\epsilon^{2}}_{x,[\hat{a}\hat{b}]}) U_{\gamma_{1}}(x,y)
\widetilde{E}^{a}(y)
U_{\gamma_{1}^{-1}}(y,x)U_{\gamma_{2}}(x,z)
\widetilde{E}^{b}(z) U_{\gamma_{2}^{-1}}(z,x)],
\end{eqnarray*}
where we have used the convention $2 \widetilde{E}^{ai}(y) \tau_{i} = 
\widetilde{E}^{a}(y)$. The expression under the trace is exactly a 
Smolin-Rovelli \cite{RoSmo_loop} loop variable 
$-T^{[ab]}(\gamma^{\epsilon^{2}}_{x,[\hat{a}\hat{b}]} \# \gamma_{xyx} \# 
\gamma_{xzx})$ based on the loop shown on Figure {\ref{loop}. 
Thus finally we get for the regulated version of $C_{I}$:

\begin{figure}
\par
\centerline{
\epsfig{figure=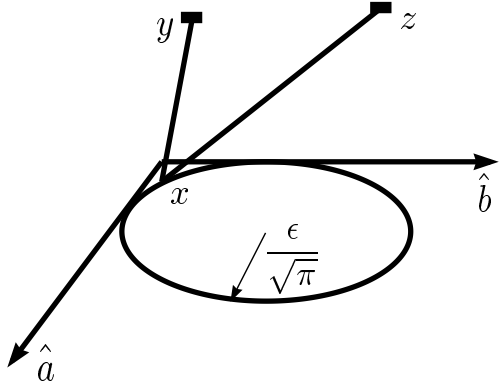,width=70mm}}
\par
\caption{Loop $\gamma^{\epsilon^{2}}_{x,[\hat{a}\hat{b}]} \# 
\gamma_{xyx} \# \gamma_{xzx}$ on which the $T$-operator is based.}
\label{loop}
\end{figure}

\[
C_{I} = \lim_{\epsilon \rightarrow 0} \lim_{\delta\rightarrow 0} 
C^{L,\delta,\epsilon}_{I},
\]
where
\begin{equation}\label{regfin}
C^{L,\delta,\epsilon}_{I} =  {L^{3} \over 2 \epsilon^{2}} 
\int_{R_{I}}{d^{3}x \over \sqrt{h(x)} } \int_{R_{I}} d^{3}y  
f_{\delta}(\tilde{x},y)\int_{R_{I}} d^{3}z  f_{\delta}(\tilde{x},z)  
\sum_{\hat{a}\hat{b}}T^{[ab]}(\gamma^{\epsilon^{2}}_{x,\hat{a}\hat{b}} 
\# \gamma_{xyx} \# \gamma_{xzx}) . 
\end{equation}
Now we promote the last expression into an operator by replacing the 
loop variable $T^{[ab]}$ with the corresponding loop operator. Thus for 
the Hamiltonian operator $\hat{H}$ we get:

\f\label{hamform}
\hat{H} = \lim_{L \rightarrow 0}{\sum_{I}^{\infty}} 
\sqrt{ -   \lim_{\epsilon \rightarrow 0} \lim_{\delta\rightarrow 0} 
\hat{C}^{L,\delta,\epsilon}_{I}
 + \Lambda \hat{V}^{2}_{I}  } ,
\ff
where $\hat{V}_{I}$ is the volume operator, as defined in \cite{DePiRo}. 
As it was shown in \cite{DePiRo} the spin network states are eigenstates 
of the volume operator so we can replace $\hat{V}_{I}$ with the 
corresponding eigenvalue for the volume of the $I$-th cube.

Thus we have a regulated version of the Hamiltonian operator with which 
we act on the spin network states.  We follow the standard procedure of 
regularization in which we apply the Hamiltonian on the states, perform 
all integrations, and at the end take the limits.

Definitions of the spin network states can be found in \cite{RoSmo2}, 
\cite{ALMMT}, \cite{Baez}, \cite{DePiRo}. For our purposes it suffices 
to recall just the basic components of the definition. The spin networks 
are defined by:

\begin{itemize}
\item a closed graph $\Gamma$ in three-space;
\item labeling of the edges by irreducible representations of $SU(2)$.
We can interpret the labels as giving the number of loop segments along
the corresponding edge;
\item intertwiners at the vertices, defining the way the loop segments 
coming from the edges are routed through the vertex.
\end{itemize}

For a rigorous description of the way the spin networks are projected on 
a plane we would need some additional details but for simplicity in our 
calculations we will assume that the spin network we are using has been 
already projected.

In our work we introduce a modification of the definition of the loop 
operators. In their standard definition  the loop operators 
$\hat{T}^{ab}[\gamma]$ are based on a loop $\gamma$ and corresponding  to 
every index there is a ``hand" attached to the loop $\gamma$. 
In our case, because the loop on which $\hat{T}^{ab}$ is based shrinks to a 
point , we have the freedom of modifying the attachment of the ``hands"  
in a way convenient for our calculations. We consider the base loop 
$\gamma$ to be a planar loop with ``hands" based on spin network edges 
of infinitesimal length $\delta$ attached to $\gamma$. These edges have 
color 2 and are denoted $\gamma_{xyx}$ and $ \gamma_{xzx}$ in equation 
Eq.\ (\ref{regfin}). It will be also convenient for us to split the points at 
which the ``hands" are attached to the loop $\gamma$ so there is a 
distance of order $\delta$ between them. This can be thought of as a
choice of decomposition of the fourvalent vertex positioned at the point 
$x$ into two trivalent vertices. 

With such a definition it can be easily shown that the standard action 
of the ``hands" is represented by connecting the spin 
network, on which $\hat{T}^{ab}$ is based, by edges of color 2 to the original 
spin network. At the place of each grasping the action creates new 
trivalent vertices. Also we multiply by a factor of $16\pi l_{Pl}^{2} 
p_{i} \Delta^{a}[e_{i}, \gamma (s)]$ whenever a ``hand" situated at 
$\gamma (s)$, corresponding to a index  ``a" of the loop operator grasps 
an edge $e_{i}$ of color $p_{i}$ from the spin network. $l_{Pl}$ is the 
Planck length and $\Delta^{a}[e_{i}, \gamma (s)]$ is the standard 
distributional expression:

\[
\Delta^{a}[e_{i}, \gamma (s)] = \oint dt \dot{e}_{i}(t)
\delta^{3}(\gamma (s), e_{i}(t) ) 
\]

The action of one of the ``hands" of the $\hat{T}^{ab}$-operator is shown on 
Figure \ref{T_act}.

\begin{figure}
\par
\centerline{
\epsfig{figure=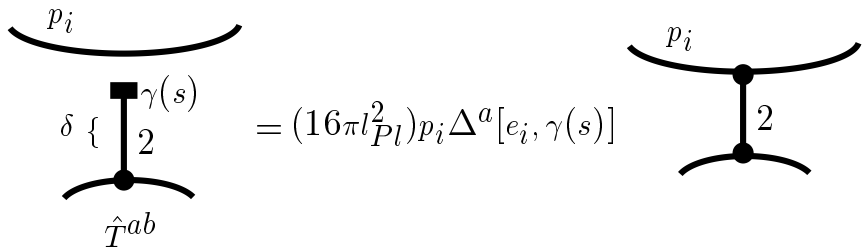,width=120mm}}
\par
\caption{Action of a ``hand" of the $\hat{T}^{ab}$-operator.}
\label{T_act}
\end{figure}

Now we are in position to apply the Hamiltonian operator to the spin 
network states. We also assume that the spin network states are normalized 
as in \cite{DePiRo}. For the action of the Hamiltonian on 
$\langle S,T|$ we get:

\begin{eqnarray}\label{act1}
\nonumber
\langle S,T|\hat{C}^{L,\delta,\epsilon}_{I} = 
 {L^{3} \over 2 \epsilon^{2}} 
\int_{R_{I}}{d^{3}x \over \sqrt{h(x)} } \int_{R_{I}} d^{3}y  
f_{\delta}(\tilde{x},y) \int_{R_{I}} d^{3}z  
f_{\delta}(\tilde{x},z) \times \\
\times \sum_{\hat{a}\hat{b}}
\{\sum_{i} \langle e_{i}|  \hat{T}^{[ab]}(\gamma^{\epsilon^{2}}_{x,\hat{a}\hat{b}} \# 
\gamma_{xyx} \# \gamma_{xzx}) 
+ \sum_{k}\sum_{i,j} \langle e_{i}^{(k)}, 
e_{j}^{(k)}|\hat{T}^{[ab]}(\gamma^{\epsilon^{2}}_{x,\hat{a}\hat{b}} 
\# \gamma_{xyx} \# \gamma_{xzx}) \}
\end{eqnarray}
In the first term of the above sum $i$ runs over all edges in the 
$I$-th cube. 
In the second term the index $k$ runs over all vertices inside the 
$I$-th cube and the indices $i$ and $j$ run over all edges joined at
the $k$-th vertex. We get two types of terms: edge terms and vertex 
terms. In an edge term the two  ``hands" of the Hamiltonian grasp one 
and the same edge of the spin network, labeled $e_{i}$, having color 
$p_{i}$. In vertex terms, the two  ``hands" grasp 
two different edges, $e_{i}^{(k)}$ and $e_{j}^{(k)}$,  joined at a 
vertex $v_{k}$. The two edges in general have different colors $p_{i}$ 
and $p_{j}$. We divide the vertex terms further into two sub-cases: 
either the tangents to the edges at the common vertex are collinear, or 
there is some angle $\theta$, different from $0^{\circ}$ and 
$180^{\circ}$ between the tangents.

We will assume that the cubes have been shrunk enough so that in a 
single cube there is at most one vertex. The evaluation of the action in 
any case can be split into two different parts -- we have an analytical 
part, coming from calculating the pre-factors in the action of the 
``hands" and evaluating the integrals, and a graphical part in which we 
complete the limiting procedure by shrinking the attached loops.

\section{Analytical action of the Hamiltonian operator on spin network 
states.}
\label{sec:analitical}
\subsection{Action on a single edge.}

For the first type of terms in the last bracket in Eq.\ (\ref{act1}) we get:

\begin{eqnarray}\label{space1}
{L^{3} \over 2 \epsilon^{2}} 
\int_{R_{I}}{d^{3}x \over \sqrt{h(x)} } \int_{R_{I}} d^{3}y  
f_{\delta}(\tilde{x},y) \int_{R_{I}} d^{3}z  f_{\delta}(\tilde{x},z)  
\sum_{\hat{a}\hat{b}}\langle e_{i}|
\hat{T}^{[ab]}(\gamma^{\epsilon^{2}}_{x,\hat{a}\hat{b}} \# \gamma_{xyx} \# 
\gamma_{xzx}) = \\
\nonumber
={L^{3} \over 2 \epsilon^{2}} (16\pi l_{Pl}^{2})^{2} p_{i}^{2} 
\int_{I} ds  \int_{I} dt 
\int_{R_{I}}{d^{3}x \over \sqrt{h(x)} }  
f_{\delta}(\tilde{x},{e_{i}}(s))   f_{\delta}(\tilde{x},{e_{i}}(t)) \times \\
\nonumber
\times \sum_{\hat{a}\hat{b}} \dot{e_{i}}^{[a}(s) \dot{e_{i}}^{b]}(t) 
\langle e_{i} \# \gamma^{\epsilon^{2}}_{x,\hat{a}\hat{b}} \# 
\gamma_{xyx} \# \gamma_{xzx}| 
\end{eqnarray}

The meaning of the notation in $\langle e_{i} \# 
\gamma^{\epsilon^{2}}_{x,\hat{a}\hat{b}} \# \gamma_{xyx} \# 
\gamma_{xzx}|$ can be understood from Figure \ref{start1}. The dashed 
circle denotes the region which will be shrunk to a point.
To proceed further we have to specify the regulating function 
$f_{\delta}(\tilde{x},y)$. We use normalized, weighted $\theta$-function:

\begin{figure}
\par
\centerline{
\epsfig{figure=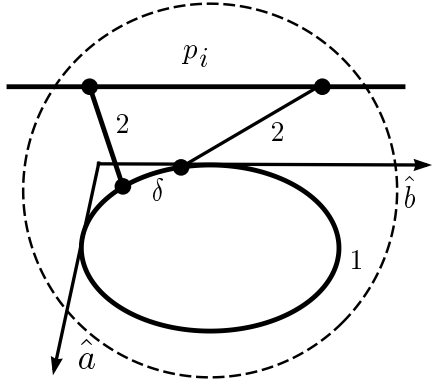,width=60mm}}
\par
\caption{Grasping of the Hamiltonian on a single edge.}
\label{start1}
\end{figure}

\begin{equation}\label{theta}
f_{\delta}(\tilde{x},y)=\sqrt{h(x)}f_{\delta}(x,y) =({3 \over 
4\pi\delta^{3}}) \sqrt{h(x)}\theta
[ \delta -|{\vec x} - { \vec y}|].
\end{equation}

After evaluation of the space integral in Eq.\ (\ref{space1}) we get :

\begin{equation}\label{Volume}
\int_{R_{I}}{d^{3}x \over \sqrt{h(x)} }f_{\delta}(\tilde{x},{e_{i}}(s))
f_{\delta}(\tilde{x},{e_{i}}(t))  ={ 3 \over 8 \pi\delta^{3}}  \left[ 2 
- { 3 \over 2 }{ d \over \delta} + \left( {d \over 2\delta} \right)^{3} 
\right] 
\end{equation}
for $d=d(s,t)=|\vec{e_{i}} (s) - \vec{e_{i}}(t)|$ less than $2\delta$ 
and zero -- otherwise.  We have assumed also that $\delta < L$.

As we are going to let $\delta$ to go to zero, this will force also the 
separation between $\vec{e_{i}}(s)$ and  $\vec{e_{i}}(t)$ to approach 
zero. This is why we keep one of the parameters, say $s$, fixed and expand 
$\vec{\dot{e_{i}}}(t)$ in power series about $s$. 
The first term in the expansion will make the whole expression vanishing, 
because of the antisymmetrization of the product 
$\dot{e_{i}}^{[a}(s) \dot{e_{i}}^{b]}(s)$. Also the distance $d$ can be 
replaced by $|(t-s)||\vec{\dot{e_{i}}}(s)|$. Combining Eq.\ (\ref{space1}) and 
Eq.\ (\ref{Volume}) we get to the lowest order in $\delta$:

\begin{eqnarray}\label{expand}
\nonumber
{L^{3} \over  \epsilon^{2} \delta^{3}}{3 \over 16\pi} (16\pi 
l_{Pl}^{2})^{2} p_{i}^{2}  
\int_{I} ds  \int dt  \left[ 2 -  3{ |(t-s)||\vec{\dot{e_{i}}}(s)|
\over 2\delta} + \left( {|(t-s)||\vec{\dot{e_{i}}}(s)| \over 2\delta} 
\right)^{3} \right] \times \\
\nonumber \times \sum_{\hat{a}\hat{b}} |t-s|\dot{e_{i}}^{[a}(s) 
\ddot{e_{i}}^{b]}(s) \langle e_{i} \# 
\gamma^{\epsilon^{2}}_{x,\hat{a}\hat{b}} \# \gamma_{xyx} \# \gamma_{xzx}|
\end{eqnarray}
The limits of integration with respect to 
$t$ are determined again by the expansion of $d(s,t)$ and are given by:

\[
t\in[  s -{2\delta\delta^{-} \over |\vec{\dot{e_{i}}}(s)|} , s + {2\delta
\delta^{+} \over |\vec{\dot{e_{i}}}(s)|}],
\]
where $\delta^{+}=1+{\cal O}(\delta)$ and $\delta^{-}=1+{\cal 
O}(\delta)$. As the integrals with respect to $s$ and $t$ are 
reparametrization invariant, we can choose 
a parametrization such that $|\vec{\dot{e_{i}}}(s)|=1$. In this 
parametrization we set:

\[ 
\vec{\dot{e_{i}}}(s) \equiv \hat{\tau}
\]
and 
\[
\vec{\ddot{e_{i}}}(s) = {\hat{n} \over \rho(s)}
\]
where $\hat{\tau}$ and $\hat{n}$ are the unit tangent and normal vectors 
to the loop and $\rho(s)$ is the curvature radius of the loop. Now we can 
perform the integration with respect to $t$ and get:

\[
\int^{s + 2\delta\delta^{+} }_{s -2\delta\delta^{-} }  dt |t-s|\left[ 2 
-  3 { |(t-s)|\over 2\delta} + \left( {|(t-s)| \over 2\delta} 
\right)^{3} \right] = 8\delta^{2}({1 \over 5} +\cal{O}(\delta)) 
\]
Thus for Eq.\ (\ref{expand}) we obtain

\[
{L^{3} \over  \epsilon^{2} \delta}{3 \over 10\pi} (16\pi l_{Pl}^{2})^{2} 
p_{i}^{2} \int_{I} {ds \over \rho(s)}  \sum_{\hat{a}\hat{b}} 
\hat{\tau}^{[a}(s) \hat{n}^{b]}(s) 
\langle e_{i} \# \gamma^{\epsilon^{2}}_{x,\hat{a}\hat{b}} \# 
\gamma_{xyx} \# \gamma_{xzx}| 
\]
Using the corresponding expression from connection representation in terms
of ho\-lo\-no\-mies it can be shown that up to terms of order $\epsilon$ the 
last sum can be written as:

\begin{equation}\label{sum}
\sum_{\hat{a}\hat{b}} \hat{\tau}^{[a}(s) \hat{n}^{b]}(s) 
\langle e_{i} \# \gamma^{\epsilon^{2}}_{x,\hat{a}\hat{b}} \# 
\gamma_{xyx} \# \gamma_{xzx}| = 
\langle e_{i} \# \gamma^{\epsilon^{2}}_{x,\hat{\tau}\hat{n}} \# 
\gamma_{xyx} \# \gamma_{xzx}| \end{equation}
where now the loop $\gamma^{\epsilon^{2}}_{x,\hat{\tau}\hat{n}} 
\# \gamma_{xyx} \# \gamma_{xzx}$ is in a plane defined by the tangent and 
the normal to the edge $e_{i}$ at the point $s$. Note that this loop is 
well defined as in the terms where the edge is a straight line and 
the normal is not defined, the curvature radius becomes infinite and 
such terms vanish. In the general case as the size $L$ of the cubes goes 
to zero we can replace the integral with respect to $s$ by its mean value:

\[
\int_{I} {ds \over \rho(s)} \langle e_{i} \# 
\gamma^{\epsilon^{2}}_{x,\hat{\tau}\hat{n}} \# \gamma_{xyx} \# 
\gamma_{xzx}|  = {\kappa L \over \rho_{I}}  \langle e_{i} \# 
\gamma^{\epsilon^{2}}_{x,\hat{\tau}\hat{n}} \# \gamma_{xyx} \# 
\gamma_{xzx}|_{I},
\]
where $\kappa$ is a number of order one, depending on the orientation
of the edge inside the cube. We finally get the analytical expression for 
the action of the Hamiltonian on a single smooth edge:

\begin{equation}\label{edge_final}
{\kappa L^{4} \over  \epsilon^{2}  \rho_{I} \delta  } {3 \over 10\pi} (16\pi 
l_{Pl}^{2})^{2} p_{i}^{2} \langle e_{i} \# 
\gamma^{\epsilon^{2}}_{x,\hat{\tau}\hat{n}} \# \gamma_{xyx} \# 
\gamma_{xzx}|_{I} 
\end{equation}
This expression is a product of a  pre-factor, containing a combination
of regulating parameters and the original spin network with attached 
additional loops, subject to graphical evaluation. The pre-factor 
will determine the way in which we take the three limits $ {\epsilon 
\rightarrow 0}$, ${\delta\rightarrow 0}$, and ${L \rightarrow 0}$ to 
make the whole expression finite.

To understand the above intermediate result we compute the action of the 
regulated diffeomorphism constraint:

\begin{equation}\label{diffeo}
\hat{C}(\vec{N}) = \lim_{L \rightarrow 0}\sum_{I} \lim_{\delta \rightarrow 0} \lim_{\epsilon \rightarrow 0} {1 \over {\epsilon}^{2}} \sum_{ab} \int_{R_{I}} d^{3}x N^{a}(x) \int_{R_{I}} d^{3}y
f_{\delta}(\tilde{x},y) \hat{T}^{b}[\gamma^{\epsilon^{2}}_{x,ab} \# 
\gamma_{xyx}] 
\end{equation}
on a single smooth edge of the spin network state. The result, up to 
finite numerical factors, is the same as Eq.\ (\ref{edge_final}) if the shift 
vector $N^{a}(x)$ is given by:

\begin{equation}\label{shift}
N^{a}(s) = \hat{n}^{a}(s){ L^{3} \over  \rho(s) \delta }.
\end{equation}

This is not surprising since a similar result is true for the calculation 
in the connection representation. To obtain this result one also must 
perform the graphical evaluation of the loop 
$\gamma^{\epsilon^{2}}_{x,\hat{\tau}\hat{n}}\#\gamma_{xyx}\#\gamma_{xzx}$ 
from Eq.\ (\ref{edge_final}) and of the loop 
$\gamma^{\epsilon^{2}}_{x,\hat{\tau}\hat{n}} \# 
\gamma_{xyx}$ from the calculation of the action of the diffeomorphism 
constraint, which shrink at the end of the calculations. We show in 
detail this evaluation later in the paper in Section 4.1. 

Thus the action of the Hamiltonian on smooth edges of the spin network 
states can be interpreted as diffeomorphism transformation in a direction 
defined by Eq.\ (\ref{shift}). This means that if we act on diffeomorphism 
invariant states, the action of the Hamiltonian on edges annihilates
them. As we assume that the states we act on a diffeomorphism invariant we 
will discard the action on smooth edges from the final result.

\subsection{Action on two edges meeting at a vertex with collinear tangents.}

\begin{figure}
\par
\centerline{
\epsfig{figure=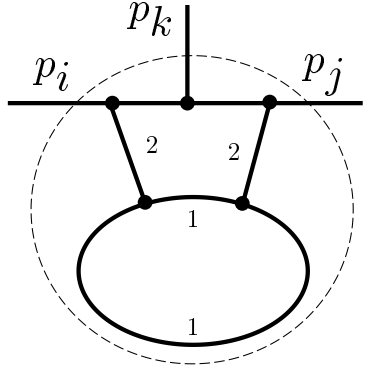,width=50mm}}
\par
\caption{Grasping of the Hamiltonian on two edges meeting at a vertex 
and having collinear tangents.}
\label{sm_vert}
\end{figure}

An example of the grasping of the Hamiltonian on two edges meeting at a 
vertex and having collinear tangents is shown on Figure \ref{sm_vert}.
The analytical calculation in this case is almost the same as with the 
case of a single edge. The only difference comes from the fact that the 
grasped edges could have different colors so instead of 
Eq.\ (\ref{edge_final}) we have:

\begin{equation}\label{2edge_final}
{\kappa L^{4} \over  \epsilon^{2}  \rho_{I} \delta  } {3 \over 10\pi} (16\pi 
l_{Pl}^{2})^{2} p_{i}p_{j} \langle (e_{i}, e_{j}) \# 
\gamma^{\epsilon^{2}}_{x,\hat{\tau}\hat{n}} \# \gamma_{xyx} \# 
\gamma_{xzx}|_{I} 
\end{equation}
In this case we can not interpret the action of the Hamiltonian as a 
diffeomorphism transformation. As discussed in \cite{Blen} in such cases 
the action can change the diffeomorphism class of the spin network. But, 
as we will see later in the paper, such terms are of order $L$ and they 
vanish after taking all limits.

\subsection{Action on two edges meeting at a vertex with non-col\-li\-ne\-ar 
tangents.}

Let the Hamiltonian grasps the edges $e_{i}$ and $e_{j}$ with colors 
$p_{i}$ and $p_{j}$ respectively. Let  $e_{i}$ and $e_{j}$ meet at some 
vertex $v_k$. We denote the part of the spin network state corresponding 
to the edges under consideration by $\langle (e_{i}, e_{j})|$. For the 
action of the Hamiltonian we have:

\begin{eqnarray*}
\langle (e_{i}, e_{j})| {L^{3} \over 2 \epsilon^{2}} 
\int_{R_{I}}{d^{3}x \over \sqrt{h(x)} } \int_{R_{I}} d^{3}y  
f_{\delta}(\tilde{x},y) \int_{R_{I}} d^{3}z  f_{\delta}(\tilde{x},z) 
\hat{T}^{[ab]}(\gamma^{\epsilon^{2}}_{x,\hat{a}\hat{b}} \# \gamma_{xyx} \# 
\gamma_{xzx}) = \\
={L^{3} \over 2 \epsilon^{2}} (16\pi l_{Pl}^{2})^{2} p_{i}p_{j}
\int_{I} ds  \int_{I} dt 
\int_{R_{I}}{d^{3}x \over \sqrt{h(x)} }  
f_{\delta}(\tilde{x},{e_{i}}(s))   f_{\delta}(\tilde{x},{e_{j}}(t)) \times \\
\times \sum_{\hat{a}\hat{b}} \dot{e_{i}}^{[a}(s) \dot{e_{j}}^{b]}(t) 
\langle (e_{i}, e_{j} ) \# \gamma^{\epsilon^{2}}_{x,\hat{a}\hat{b}} \# 
\gamma_{xyx} \# \gamma_{xzx}| \end{eqnarray*}
Again $\langle (e_{i}, e_{j} ) \# 
\gamma^{\epsilon^{2}}_{x,\hat{a}\hat{b}} \# \gamma_{xyx} \# 
\gamma_{xzx}|$ denotes the action of grasping of the Hamiltonian. Using 
Eq.\ (\ref{Volume}) we can evaluate the spatial integral to get

\begin{eqnarray}\label{vert}
\nonumber
{L^{3} \over  \epsilon^{2} \delta^{3}}{3 \over 16\pi} (16\pi 
l_{Pl}^{2})^{2} p_{i} p_{j} \int_{I} ds  \int_{I}  dt \left[ 2 - 3 { 
|\vec{e_{i}}(s) - \vec{e_{j}}(t)| \over 2\delta} + \left( 
{|\vec{e_{i}}(s) - \vec{e_{j}}(t)| \over 2\delta} \right)^{3} \right]
\times \\
\times\sum_{\hat{a}\hat{b}} \dot{e_{i}}^{[a}(s) \dot{e_{j}}^{b]}(t) 
\langle (e_{i}, e_{j} ) \# \gamma^{\epsilon^{2}}_{x,\hat{a}\hat{b}} \# 
\gamma_{xyx} \# \gamma_{xzx}|_{I}. 
\end{eqnarray}

We use a parametrization such that $ |\dot{\vec{e_{i}}}| = 
|\dot{\vec{e_{j}}}| = 1 $ so the distance $d(s,t) = |\vec{e_{i}}(s) - 
\vec{e_{j}}(t)| $ becomes simply $d(s,t) = |s-t|$. 
Also let $\eta$ and $\zeta$ be the unit tangent vectors at the vertex 
and let the angle between them be $\theta_{ij}$. Performing the integration 
with respect to $s$ and $t$ we get for Eq.\ (\ref{vert}):

\[
{L^{3} \over  \epsilon^{2} \delta}{3\theta_{ij} \over 10 \pi \sin{\theta}} 
(16\pi l_{Pl}^{2})^{2} p_{i} p_{j}   
\sum_{\hat{a}\hat{b}} \hat{\eta}^{[a} \hat{\zeta}^{b]} \langle (e_{i}, 
e_{j} ) \# \gamma^{\epsilon^{2}}_{x,\hat{a}\hat{b}} \# \gamma_{xyx} \# 
\gamma_{xzx}|_{I}. \]
Again we can use a formula which can be proven by using the 
corresponding expression in the connection representation, namely:

\[
\sum_{\hat{a}\hat{b}} \hat{\eta}^{[a} \hat{\zeta}^{b]} \langle (e_{i}, 
e_{j} ) \# \gamma^{\epsilon^{2}}_{x,\hat{a}\hat{b}}  \# \gamma_{xyx} \# 
\gamma_{xzx}|_{I} = \sin{\theta} \langle (e_{i}, e_{j} ) \# 
\gamma^{\epsilon^{2}}_{x,\hat{\eta}\hat{\zeta}}\# \gamma_{xyx} \# 
\gamma_{xzx}|_{I}, \]
to transform the sum in Eq.\ (\ref{vert}). In the last expression 
$\gamma^{\epsilon^{2}}_{x,\hat{\eta}\hat{\zeta}}$ is a loop in the plane 
defined by the two tangent vectors $\hat{\eta}$ and $\hat{\zeta}$. 
Thus finally we get for the action of the Hamiltonian on a vertex:

\begin{equation}\label{vertex_final}
{L^{3} \over  \epsilon^{2} \delta}{3\theta_{ij}  \over 10 \pi } (16\pi 
l_{Pl}^{2})^{2} p_{i} p_{j}   \langle (e_{i}, e_{j} ) \# 
\gamma^{\epsilon^{2}}_{x,\hat{\eta}\hat{\zeta}} \# \gamma_{xyx} \# 
\gamma_{xzx}|_{I}. 
\end{equation}

In the obtained expression we can separate between the pre-factor and the
loop-deformed spin network state. The pre-factor contains a combination 
of the regulating parameters and an implicit dependence on the arbitrary 
background metric through the angle $\theta_{ij}$.
The loop deformation will be subject to a graphical evaluation later in 
the paper.

\subsection{Taking the limits.}
Now we are in position to take the limits in the computation. Basically
the limits appear in our calculation in two different ways. First the 
pre-factors in Eq.\ (\ref{2edge_final})  and Eq.\ (\ref{vertex_final}) contain 
combinations of the regulating parameters. The limits should be taken in 
such a way so these combinations yield finite results. Second, the loop 
on which the loop operator $\hat{T}^{ab}$ is based shrinks 
to a point, together with its ``hands" and this leads to the graphical 
evaluation. 

The combinations of parameters are:

\begin{eqnarray}
{\kappa L^{4} \over  \epsilon^{2}  \delta  } & \ 
{\rm and}\  & {L^{3} \over  \epsilon^{2} \delta}
\end{eqnarray}
for Eq.\ (\ref{2edge_final})  and Eq.\ (\ref{vertex_final}) respectively. 
To insure finiteness of our expressions let us take the three 
limits $ {\epsilon \rightarrow 0}$, ${\delta\rightarrow 0}$, and 
${L \rightarrow 0}$ along a plane in the $(L, \epsilon, \delta)$ 
parameter space, defined by:

\[
{L^{3} \over  \epsilon^{2} \delta}  = Z
\]
where $Z$ is an arbitrary constant, chosen in such a way so the relative 
order of taking the limits is satisfied. This order is determined by the 
conditions that $\delta < L$ and $\delta \ll \epsilon $. We have to be 
careful about the way the limits in $L$ and $\delta$ are taken, as the 
diffeomorphism vector Eq.\ (\ref{shift}) contains the ratio $L^{3}/\delta$, 
so we would not like $\delta$ to go to zero much faster than $L$. If we 
set $L^{3}/\delta = {\rm const}$ then this will be enough to prevent the 
diffeomorphism from becoming infinite. 

We can now write the general formula for the analytical action of the 
Hamiltonian on diffeomorphism invariant spin network states. All the 
cubes which are empty give zero. 

The result from the action on different edges meeting at a vertex with 
collinear tangents Eq.\ (\ref{2edge_final}) can be written as :

\[
{3\kappa Z L  \over 10\pi\rho_{I}} (16\pi l_{Pl}^{2})^{2} p_{i}p_{j}  
\langle (e_{i},e_{j}) \# \gamma^{\epsilon^{2}}_{x,\hat{\tau}\hat{n}} \# 
\gamma_{xyx} \# \gamma_{xzx}|_{I} 
\]
and vanishes as $L$ goes to zero.

Adding  up all contributions from the vertices with non-collinear 
tangents we get: 

\begin{equation}\label{act_before}
\langle S,T| \hat{C}^{L,\delta,\epsilon}_{I} = 
{3 Z \over 10 \pi} (16\pi l_{Pl}^{2})^{2}
\sum_{i,j}  p_{i} p_{j}  \theta_{ij}\overline{ \langle 
(e_{i}^{(k)}, e_{j}^{(k)} ) \# 
\gamma^{\epsilon^{2}}_{x,\hat{\eta}\hat{\zeta}}\# \gamma_{xyx} \# 
\gamma_{xzx}|} \end{equation}
In the above sums the indices $i$ and $j$ run over all edges joined
at the vertex $v_k$ which is in the $I$-th cube. The sum  contains the 
arbitrary but finite constant $Z$ and the angle $\theta_{ij}$ between 
the edges at the vertex $v_k$. The bar over the state represents the 
fact that we still have to perform the shrinking of the attached loops 
and thus to evaluate the graph. 

We get overall action which is finite but background dependent because 
of the presence of the angle $\theta_{ij}$. This is a problem which can be 
solved by redefining the way we approximate the curvature $F_{ab}$. Let 
the holonomy which approximates the curvature be based on a loop with area 
$\theta_{ij} \epsilon^{2}$ instead of $\epsilon^{2}$. Then all the 
calculations go through but at the end there is an extra factor of 
$\theta_{ij}$ in the denominator of Eq.\ (\ref{act_before}) to cancel 
the corresponding factor from the numerator. It is important to notice 
that although the explicit inclusion of the factor of $\theta_{ij}$ formally 
solves the problem of background dependence, the situation is not 
completely satisfactory.  The computation of the evolution of 
the spin network states involves at each step measuring the angle 
$\theta_{ij}$ between each pair of tangent vectors. Technically this might 
require the introduction in the definition of the Hamiltonian of an operator, 
measuring the angle $\theta_{ij}$. This issue requires
further investigation.

\section{Graphical action of the Hamiltonian.}
\label{sec:graph}

To evaluate the deformed spin network state from equation 
Eq.\ (\ref{act_before}) we use some techniques from the recoupling theory of 
colored trivalent links and knots. First let us 
consider a general situation in which a loop attached to a spin network 
is shrunk to a point. An important observation is that the evaluation 
is local, in the sense that whatever result we obtain, it is independent of 
the way the spin network is connected  outside of the circle denoting 
the shrinking region.

To evaluate the attached loop we have to do the following

\begin{itemize}
\item Expand all the edges which are entirely in the shrinking region as an 
antisymmetrized sum with all possible crossings between the loop segments.
\item Resolve each crossing of single loop segments in the shrinking region
according to the binor Mandelstam identity (see \cite{DePiRo} for discussion).
\item Associate to each closed single loop the ``loop value" (-2).
\item Smoothen the obtained graph to get if possible the original spin 
network. 
\end{itemize}

But these are exactly the operations which occur in the evaluation of the
Kauffman bracket \cite{KaLi} . The Kauffman bracket is an invariant of 
regular isotopy of colored knots and links with trivalent vertices. Thus 
for the evaluation of the graphical action of the loop operators we can 
use the techniques of recoupling theory, developed for computing the 
Kauffman bracket \cite{DePiRo,KaLi,MaSmo,BoMaSmo}.
What makes these techniques 
powerful is the fact that we do the computations blockwise. First we 
work out the analytical expressions for the most simple graphs. Then in 
more complicated calculations we identify the simple graphs and replace 
them with the corresponding analytical expressions. This is possible 
because the evaluation of the graphs is local -- we 
can shrink to a point just a portion of the whole graph, keeping the 
remaining parts fixed. The basic formulae of the recoupling theory are 
summarized in the Appendix. 
With the use of these formulae we compute the graphical action of the 
Hamiltonian on smooth edges and on vertices.

\subsection{Evaluation of the graphical action for edges.}

We perform the graphical evaluation for the grasp of a smooth 
edge to show that the graphical action of the Hamiltonian on smooth edges 
is equivalent to diffeomorphism transformation. We start from the expression  
$\langle 
e_{i} \# \gamma^{\epsilon^{2}}_{x,\hat{\tau}\hat{n}} \# \gamma_{xyx} \# 
\gamma_{xzx}|$. The grasping is shown on Figure \ref{start1}.
First we use the basic recoupling formula Eq.\ (\ref{recop}) from the Appendix 
on one of the edges. As a result  a new link of color $n$ appears in the 
new configuration. The allowed values for  $n$ are determined by the 
basic properties of the trivalent vertices -- the only values it can 
take are $p_i \pm 1$. 

\[
\langle e_{i} \# \gamma^{\epsilon^{2}}_{x,\hat{\tau}\hat{n}} \# 
\gamma_{xyx} \# \gamma_{xzx}| = \kd{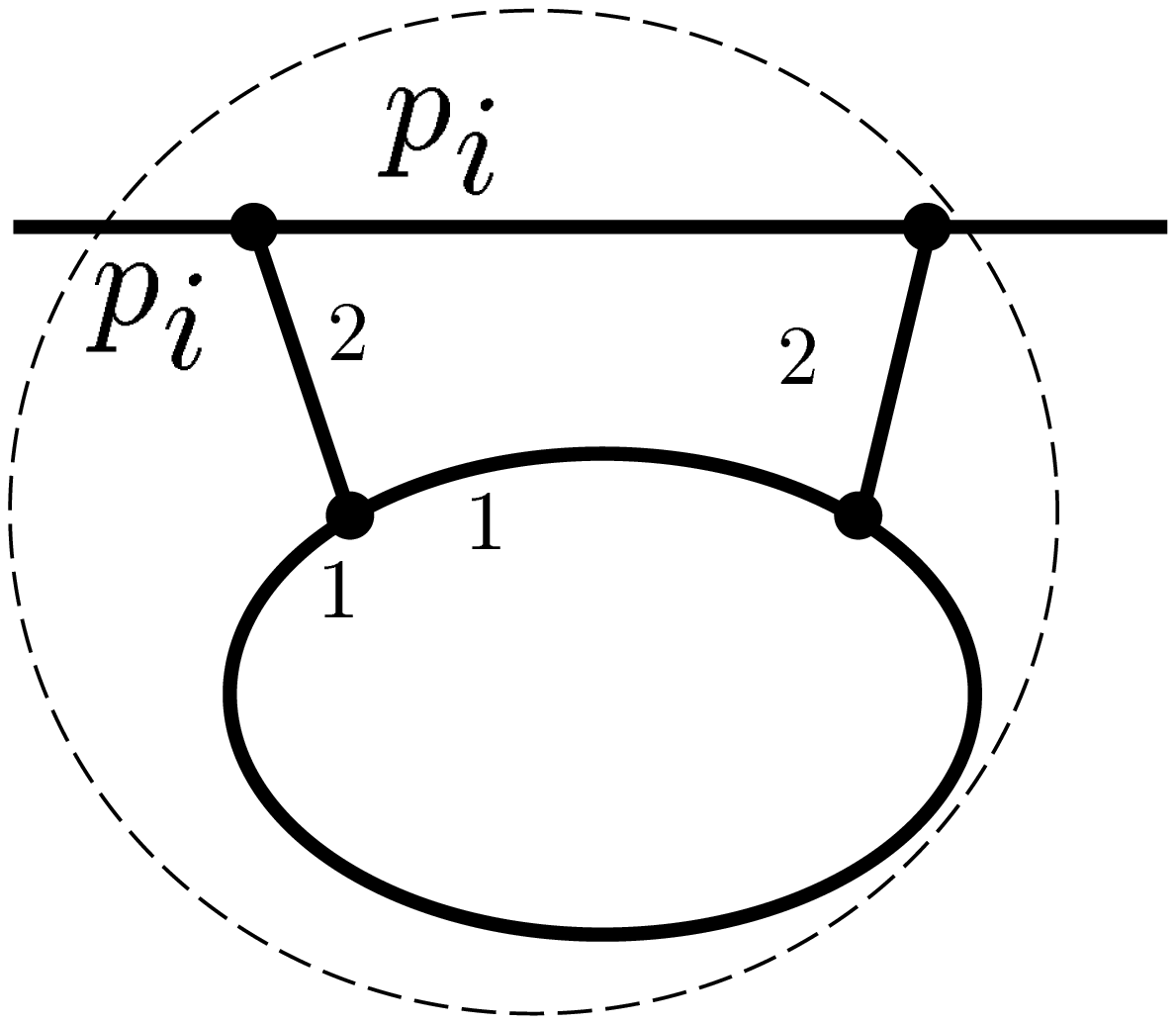} = \sum_{n=p_i \pm 1}
\left\{ \begin{array}{ccc} p_i & p_i & n \\ 1 & 1 & 2
\end{array} \right\} \kd{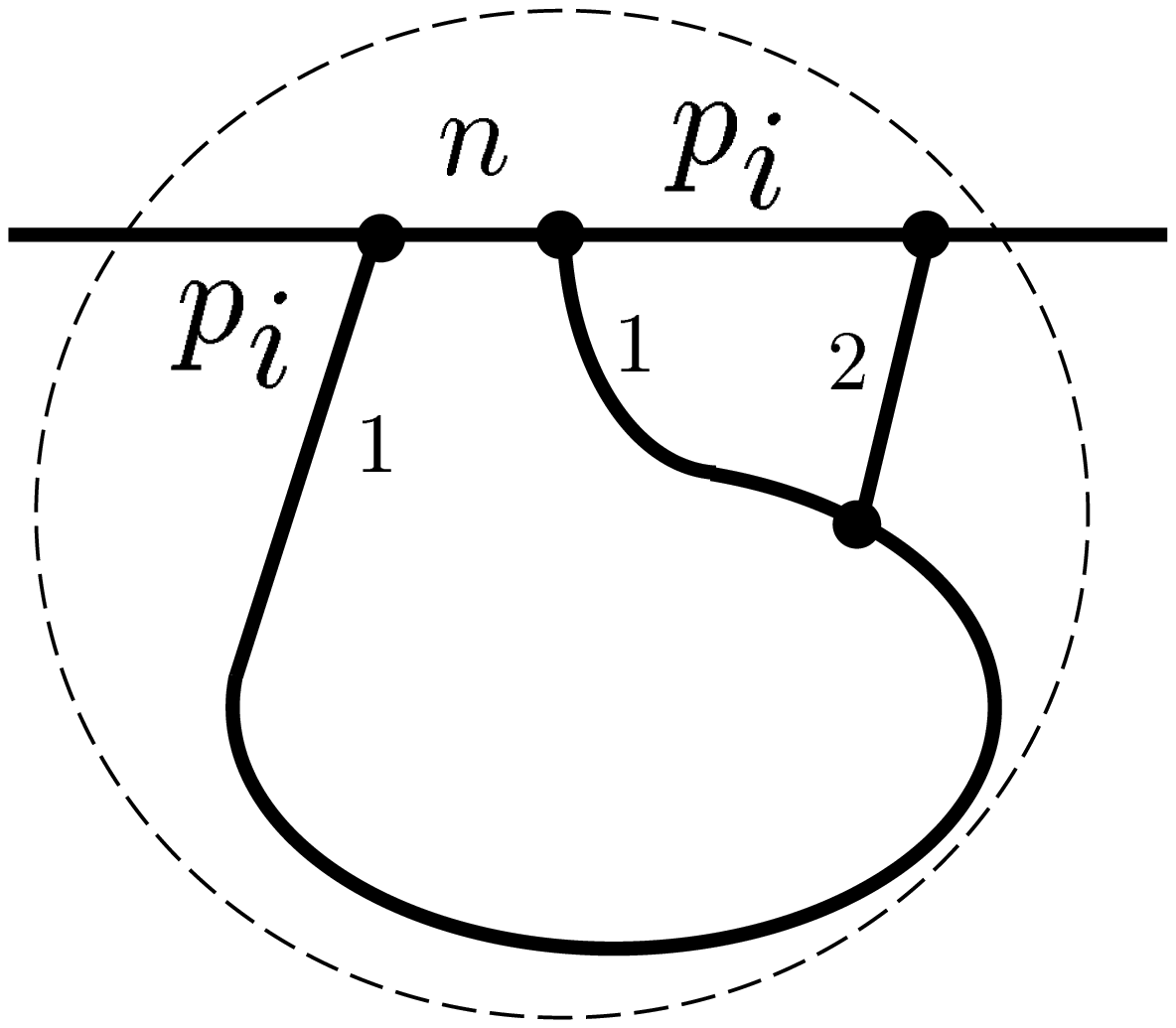} 
\]
On the right hand side appears the 6j-symbol, as defined in \cite{KaLi}. 
We repeat the same step with the other ``hand" and get:

\[
\sum_{n=p_i \pm 1} \sum_{m=p_i \pm 1}
\left\{ \begin{array}{ccc} p_i & p_i & n \\ 1 & 1 & 2
\end{array} \right\} \left\{ \begin{array}{ccc} p_i & p_i & m \\ 1 & 1 & 2
\end{array} \right\} \kd{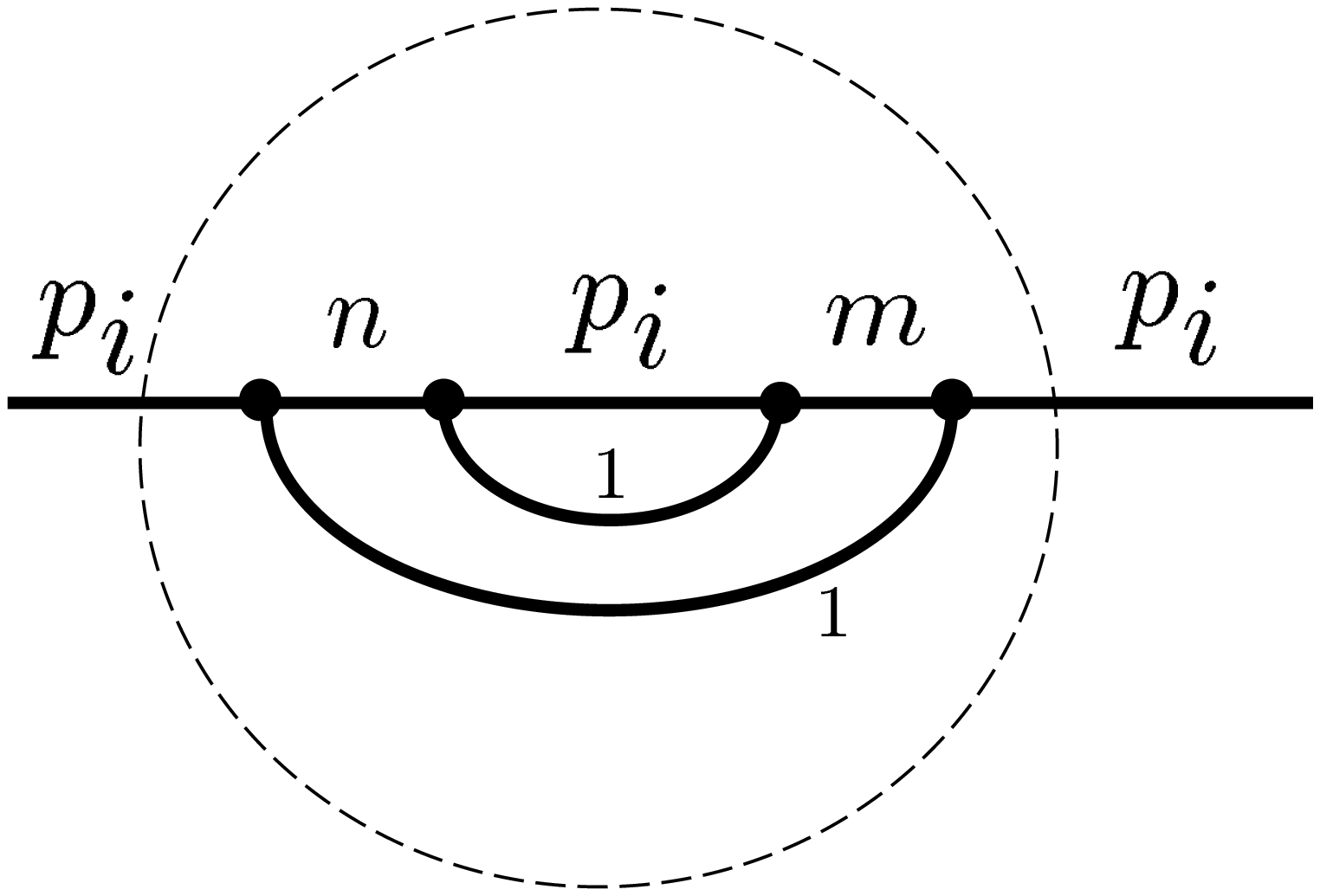} 
\]

As a next step we remove the internal ``bubble" using the 
$\vartheta$-net from equation Eq.\ (\ref{theta_net}) from the Appendix to 
obtain:

\[
\langle e_{i} \# \gamma^{\epsilon^{2}}_{x,\hat{\tau}\hat{n}} \# \gamma_{xyx} 
\# \gamma_{xzx}| =  \sum_{n=p_i \pm 1} (-1)^{n}
\left\{ \begin{array}{ccc} p_i & p_i & n \\ 1 & 1 & 2
\end{array} \right\}^{2} {\vartheta(p_i, 1, n) \over n+1}\kd{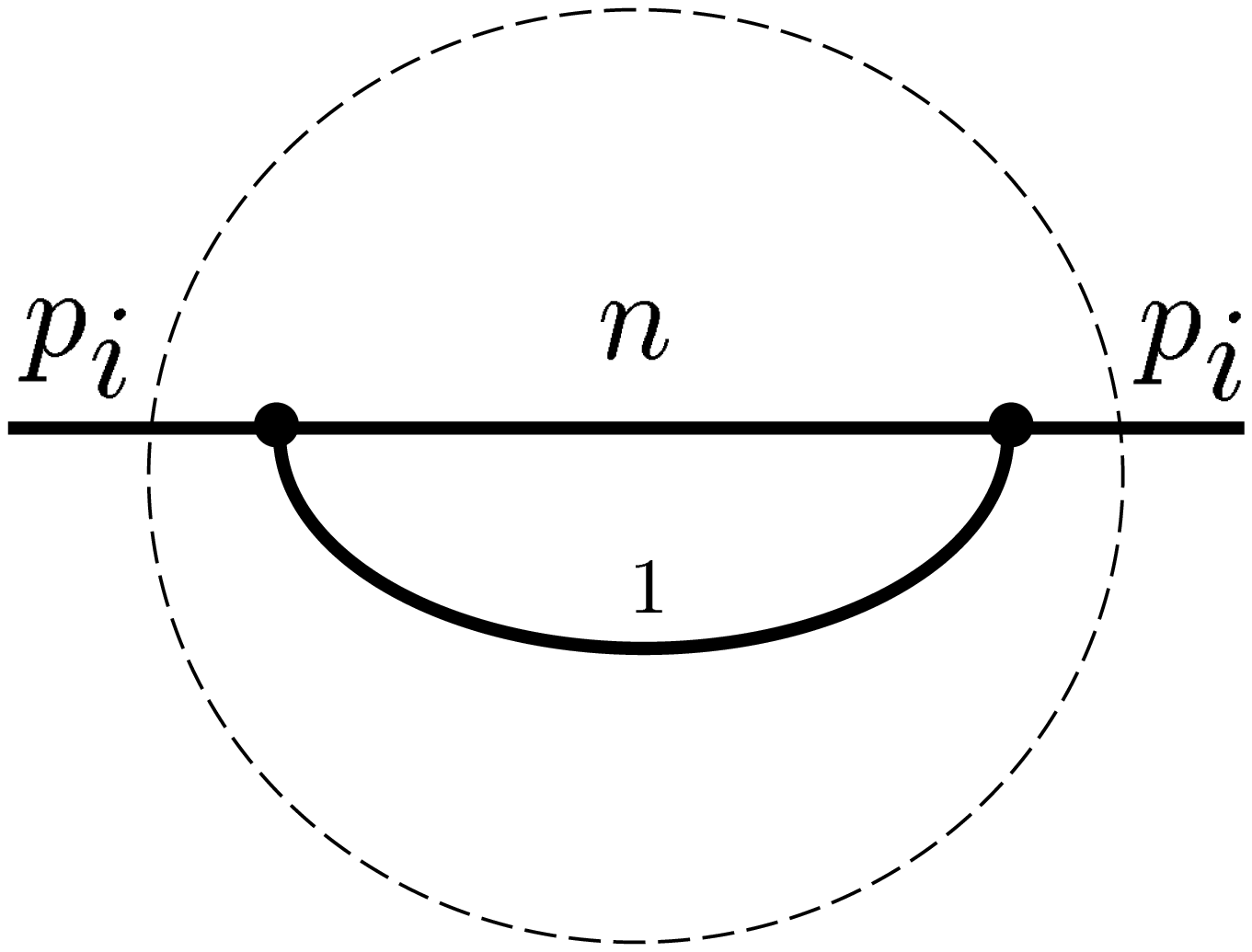}
\]

To continue further we have to make more careful analysis of the way the 
limit is taken, when the states are diffeomorphism invariant. As this is 
not relevant for our considerations we will stop here and show that the 
above expression is analogous to the result from the 
action of the diffeomorphism constraint. To this aim we act with 
Eq.\ (\ref{diffeo}) on a smooth edge. There the loop deformation is given by 
the expression $\gamma^{\epsilon^{2}}_{x,\hat{\tau}\hat{n}} \# 
\gamma_{xyx}$. To evaluate graphically this expression we use the basic 
recoupling formula Eq.\ (\ref{recop}) and immediately get:

\[
\langle e_{i} \# \gamma^{\epsilon^{2}}_{x,\hat{\tau}\hat{n}} \# \gamma_{xyx}| = 
\kd{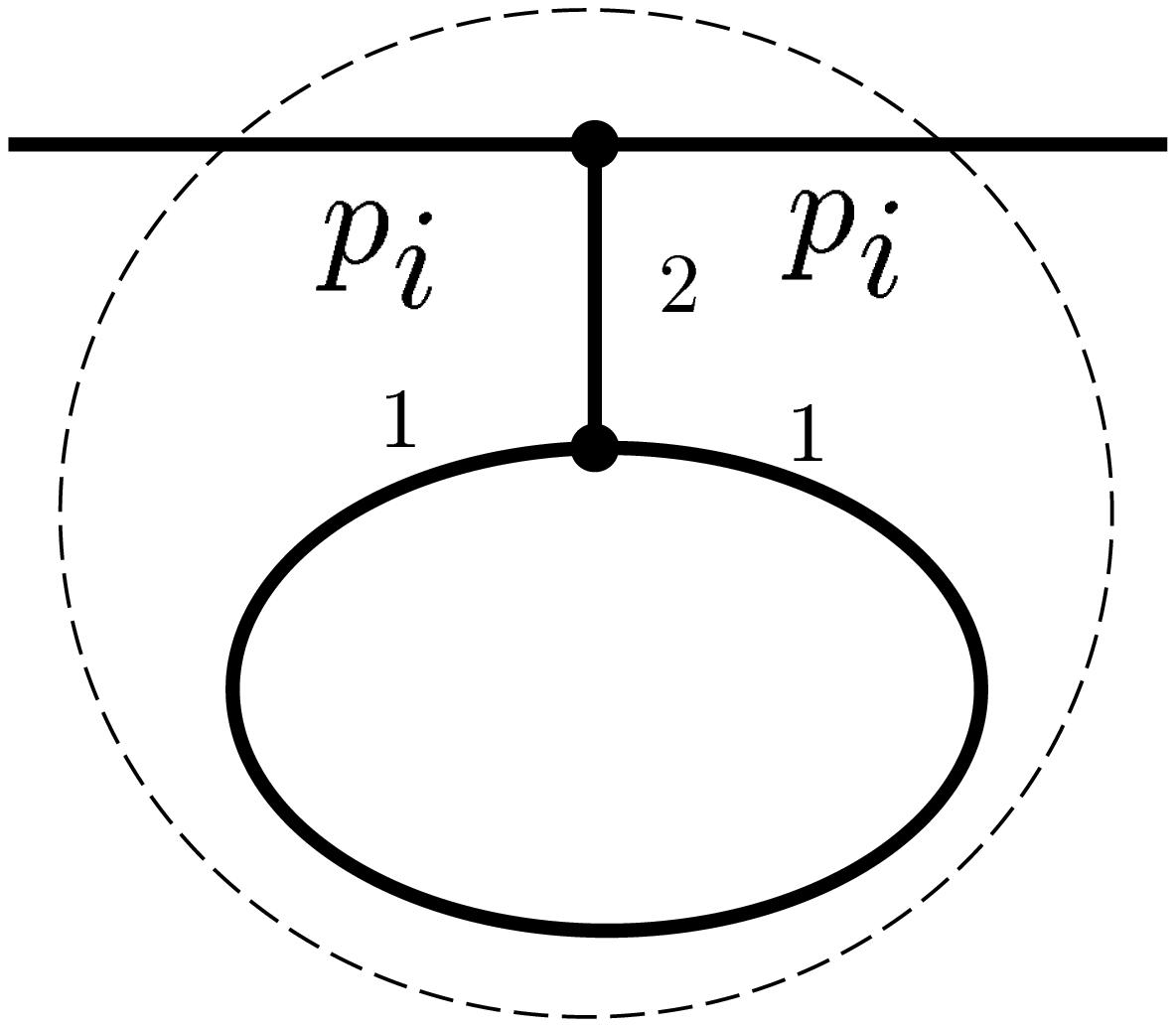} = \sum_{n=p_i \pm 1}
\left\{ \begin{array}{ccc} p_i & p_i & n \\ 1 & 1 & 2
\end{array} \right\} \kd{recoup4} 
\]

Thus we see that in both cases of the action of the Hamiltonian and of the 
diffeomorphism constraints, the graphical evaluation leads to the a 
``bubble" on the smooth edge with different 
numerical factors. This proves that the two graphical actions are equivalent 
and thus we can disregard all terms in the action of the Hamiltonian on 
smooth edges of diffeomorphism invariant states.

\subsection{Evaluation of the graphical action for vertices.}

We apply the techniques from the recoupling theory of colored graphs 
also to evaluate the action of the Hamiltonian when it grasps two edges 
meeting at a vertex. For simplicity we consider only the action on one 
pair of edges joint at a trivalent vertex. We have initially the 
expression $ \langle (e_{i}, e_{j} ) \# 
\gamma^{\epsilon^{2}}_{x,\hat{\eta}\hat{\zeta}} \# \gamma_{xyx} \# 
\gamma_{xzx}|$ which can be transformed with the use of the recoupling 
formula Eq.\ (\ref{recop}), given in the Appendix. We obtain the following 
result:

\[
\langle (e_{i}, e_{j} ) \# 
\gamma^{\epsilon^{2}}_{x,\hat{\eta}\hat{\zeta}} \# 
\gamma_{xyx} \# \gamma_{xzx}| = \kd{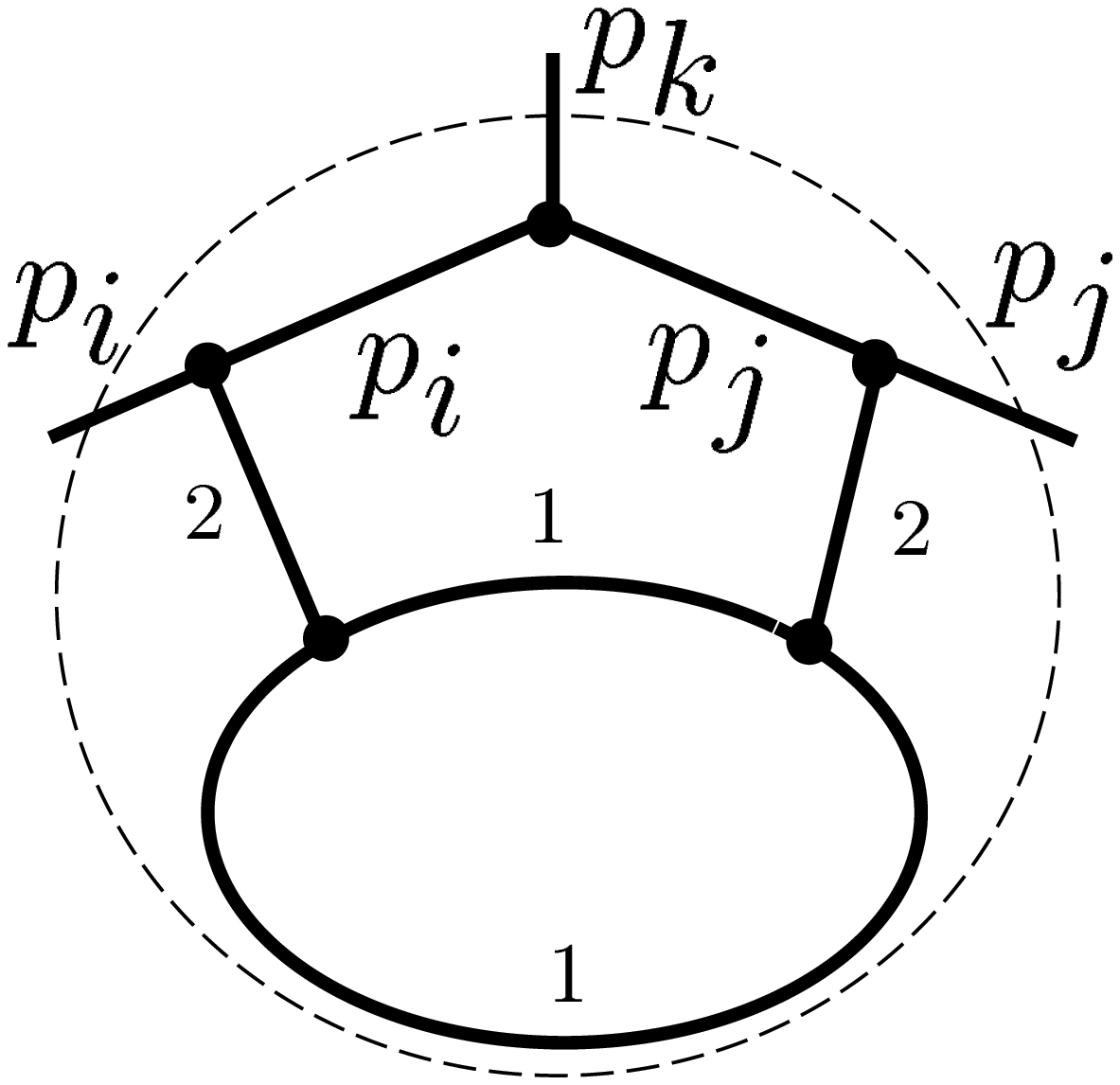} = \sum_{n=p_i \pm 1}
\left\{ \begin{array}{ccc} p_i & p_i & n \\ 1 & 1 & 2
\end{array} \right\} \kd{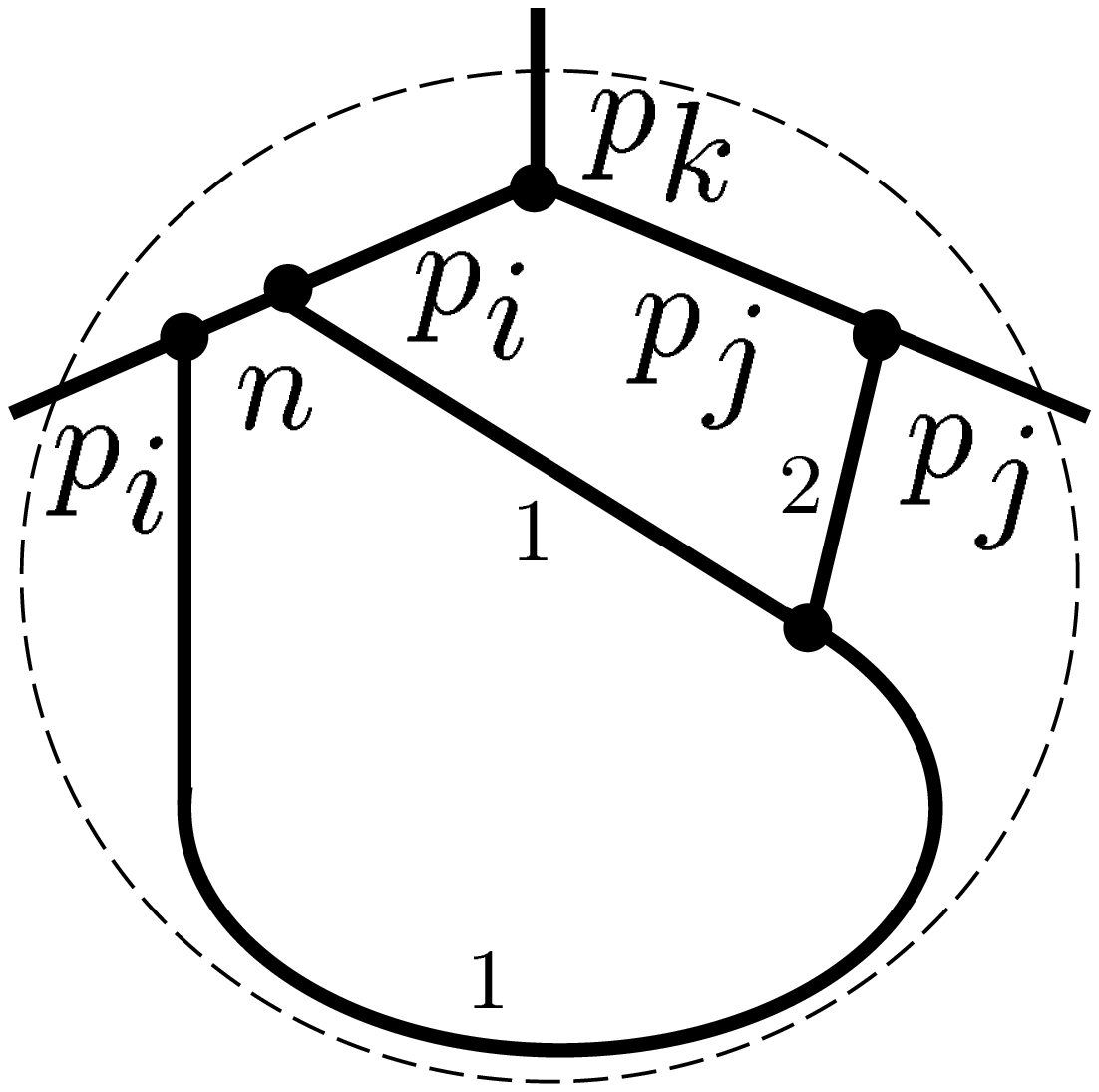}. 
\]
Then we repeat the same step with the other ``hand" of the $T$-operator 
to get:

\[
\sum_{n=p_i \pm 1} \sum_{m=p_j \pm 1}
\left\{ \begin{array}{ccc} p_i & p_i & n \\ 1 & 1 & 2
\end{array} \right\} \left\{ \begin{array}{ccc} p_j & p_j & m \\ 1 & 1 & 2
\end{array} \right\} \kd{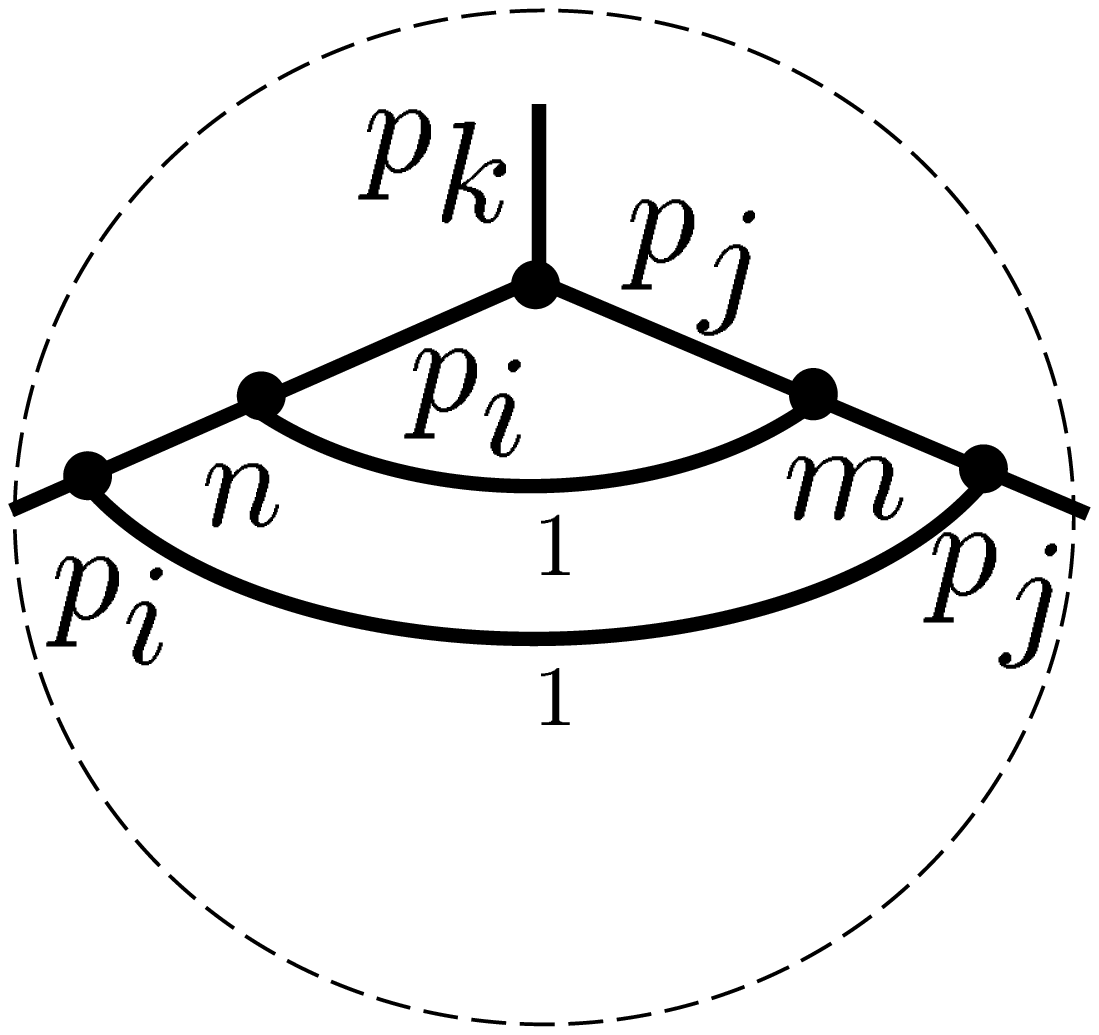}. 
\]

The the inner triangular diagram of the above graph can be evaluated 
with the use of one recoupling formula and one ``bubble" removal (see 
the Appendix):

\begin{eqnarray*}
\langle (e_{i}, e_{j} ) \# \gamma^{\epsilon^{2}}_{x,\hat{\eta}\hat{\zeta}} 
\# \gamma_{xyx} \# \gamma_{xzx}| = \sum_{n=p_i \pm 1} \sum_{m=p_j \pm 1}
\left\{ \begin{array}{ccc} p_i & p_i & n \\ 1 & 1 & 2
\end{array} \right\} \left\{ \begin{array}{ccc} p_j & p_j & m \\ 1 & 1 & 2
\end{array} \right\} \times \\
\times  \left\{ \begin{array}{ccc} n & p_i & p_k \\ p_j & m & 1
\end{array} \right\} { (-1)^{p_{k}} \vartheta(p_i , p_j , p_k ) \over 
p_{k}+1} \kd{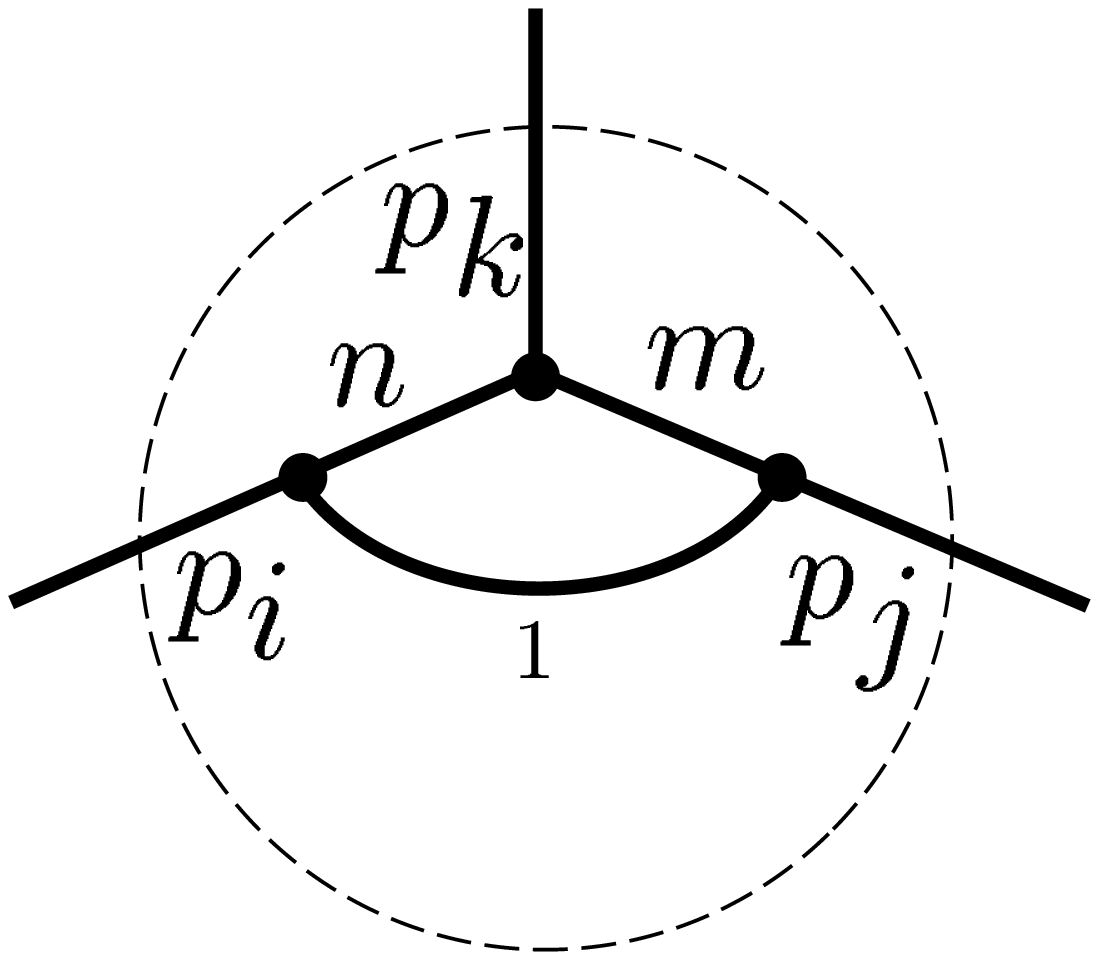}. 
\end{eqnarray*}

The last step in the graphical calculation requires more careful 
consideration. As discussed in \cite{RoSmo1} when we act on 
diffeomorphism invariant states,
the spin network remains in its knot class, although the area of the 
last remaining  loop goes to zero. To understand this situation let us 
discuss the action of the Hamiltonian from the point of view of 
diffeomorphism invariance. We start with a state which is diffeomorphism 
class of spin network and a Hamiltonian, also based on  a class of 
diffeomorphism invariant smooth loops with ``hands". To perform the
action we introduce a background metric, break the diffeomorphism 
invariance, and introduce notions of lengths and areas. It is in this 
non-invariant sense in which we can talk about ``loops of some area 
shrinking to a fixed point" and about ``hands of infinitesimal length". 
Also all the formulae we apply from the recoupling theory are
based on the ideas of replacing loops by their loop values and 
recoupling, which are diffeomorphically non-invariant operations. We assume
that we can perform these operations as far as we recover at the end the 
``right" diffeomorphism class. What we expect at the end is the original 
spin network attached through two ``hands" to a loop
which in the non-diffeomorphism limit shrinks to a vertex. 

We can choose the attached loop in different way but we want our choice 
to be consistent with the operations we perform in a non-invariant 
fashion. One possible such choice  is the attached loop to 
connect two edges joined at a vertex and then to run along the edges of 
the original spin network. This means that when the state we are acting 
on is diffeomorphism invariant taking the last limit is trivial -- the 
graph does not change. Thus finally we get:

\begin{eqnarray}\label{act_after}
\langle S,T| \hat{C}^{L,\delta,\epsilon}_{I} = 
{3 Z \over 10 \pi} (16\pi l_{Pl}^{2})^{2}
\sum_{i,j}  p_{i} p_{j} \sum_{n=p_i \pm 1} \sum_{m=p_j \pm 1}
\left\{ \begin{array}{ccc} p_i & p_i & n \\ 1 & 1 & 2
\end{array} \right\} \times \\
\nonumber
\times \left\{ \begin{array}{ccc} p_j & p_j & m \\ 1 & 1 & 2
\end{array} \right\} \left\{ \begin{array}{ccc} n & p_i & p_k \\ p_j & m & 1
\end{array} \right\} {(-1)^{p_{k}}\vartheta(p_i , p_j , p_k ) \over 
p_{k}+1} \kd{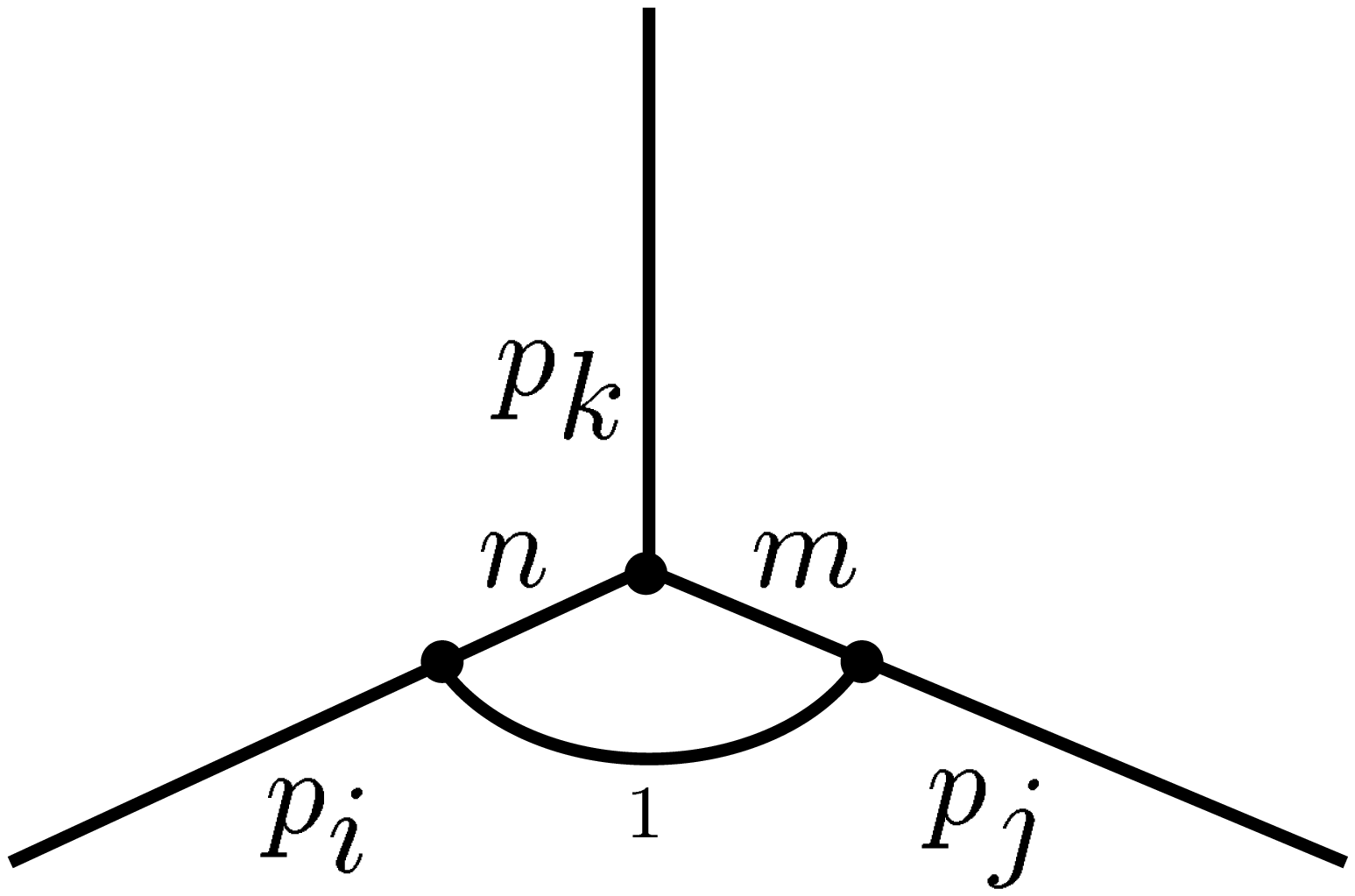}, 
\end{eqnarray}
where the vertex having the additinal edge attached to it is in the 
$I$-th cube.
The state $\langle S,T| $ we acted on is normalized
according to the normalization introduced in \cite{DePiRo}. If
the final states are also to be normalized we have to introduce some
additional factors in Eq.\ (\ref{act_after}). As the old trivalent vertex 
has been transformed into a new one, two new edges and two new vertices 
have been added we get for the final normalized sum of states:

\begin{eqnarray}\label{final_norm}
\langle S,T|\hat{C}^{L,\delta,\epsilon}_{I} = {3 Z(16\pi 
l_{Pl}^{2})^{2} \over 10 \pi} \sum_{i,j}  p_{i} p_{j} 
\sum_{n=p_i \pm 1} \sum_{m=p_j \pm 1} \sqrt{{-2 \Delta_{n}\Delta_{m}
\vartheta(p_i, p_j, p_k) \over \vartheta(1, p_i, n) \vartheta (1, p_j, m) 
\vartheta(n, m, p_k)}}
\times \\
\nonumber
\times \left\{ \begin{array}{ccc} p_i & p_i & n \\ 1 & 1 & 2
\end{array} \right\} \left\{ \begin{array}{ccc} p_j & p_j & m \\ 1 & 1 & 2
\end{array} \right\} \left\{ \begin{array}{ccc} n & p_i & p_k \\ p_j & m & 1
\end{array} \right\} {(-1)^{p_{k}}\vartheta(p_i , p_j , p_k ) \over 
p_{k}+1} \kd{vertex4}. 
\end{eqnarray}

The action of the Hamiltonian constraint $\hat{C}$ as part of the physical
Hamiltonian $\hat{H}$ can be described as follows:
When the Hamiltonian constraint $\hat{C}$ acts on edges, the action can be 
interpreted as a diffeomorphism transformation. Thus the Hamiltonian 
constraint acts non-trivially on diffeomorphism invariant states only when it
acts on the vertices of the spin network. The action on vertices amounts to 
adding a new edge attached through new trivalent vertices  to each pair 
of edges at each vertex. Each new edge has color 1.  Also the colors of 
the edges connecting the new vertices and the original one change. This 
change can be explained with the requirement that
the new graph is again a spin network. The obtained states are 
multiplied by finite factors which, although cumbersome, are 
straightforward to compute.

\section{The Hamiltonian eigenspectrum}
\label{sec:spec}
Let us recall that the Hamiltonian defined through the ``clock'' field 
has the form Eq.\ (\ref{hamform}):

\f\label{hamform2}
\hat{H} = \lim_{L \rightarrow 0}{\sum_{I}^{\infty}} 
\sqrt{ -   \lim_{\epsilon \rightarrow 0} \lim_{\delta\rightarrow 0} 
\hat{C}^{L,\delta,\epsilon}_{I}
 + \Lambda \hat{V}^{2}_{I}  }.
\ff
To find the eigenspectrum of the Hamiltonian operator we have to handle in
a satisfactory manner the square root in Eq.\ (\ref{hamform2}). Since we have 
not been able to complete this task we will only describe the directions 
for future work.
 
Since both the Hamiltonian constraint and the volume operator in 
Eq.\ (\ref{hamform2}) give 
non-zero results only when acting on vertices, the above sum reduces to a sum 
only over the cubes in which there is a vertex. Thus the sum in 
Eq.\ (\ref{hamform2}) becomes finite. Also all the terms coming from the action 
on different vertices are independent from each other.
This justifies the following strategy for computing the 
Hamiltonian eigenspectrum: We can use the result from the action of the
	expression under the square root on vertices -- the action of the first 
term is given by Eq.\ (\ref{final_norm}) and the action of the volume piece 
have been computed for example in \cite{DePiRo}. To proceed further we might 
have to distinguish between the action on bivalent and trivalent vertices 
on one hand, as having zero volume, and the action on higher-valence 
vertices. In any case we will have to diagonalize the expression 
Eq.\ (\ref{final_norm}). The problem is that {\it a priori} we can not expect 
the matrix defined by
Eq.\ (\ref{final_norm}) to be diagonalizable. Since the expression 
for the Hamiltonian constraint we start with is a real one we can 
assume that there exists an appropriate symmetrization which can make the 
constraint operator self-adjoint and its matrx symmetric. This is why 
to make Eq.\ (\ref{final_norm}) diagonalizable we could replace it
by the semi-sum of it and its transposed. Then we will have to diagonalize 
the sum of the Hamiltonian and volume matrices and to take the square root.
The result will be an infinite dimmensional diagonal matrix with the
Hamiltonian eigenvalues as its elements.

\section{Conclusion}
\label{sec:concl}
The action of the Hamiltonian constraint on a spin network state can 
be described as a transition 
of the spin networks on which the state is based from  one knot class of 
graphs with vertices into another one.  In this transition the coloring 
of certain edges changes so the requirements for the graph to be a spin 
network remain satisfied.  At the same time 
each new state is multiplied by a factor, which carries the appropriate 
dimensions, contains an arbitrary constant, and also includes the result 
from the computation in recoupling theory. 

In the paper we considered in detail only trivalent vertices  but the 
generalization to higher valence vertices is straightforward. As discussed in 
\cite{RoSmo0},  \cite{DePiRo}, any higher-valence vertex can be decomposed 
into a set of infinitesimally displaced trivalent ones. The analytical 
calculation we performed did not depend to the valence of the vertex. The 
difference with the considered case will appear in the graphical 
evaluation where we again can apply the recoupling theory. It can be shown 
that as a result in the same way as with 
the trivalent vertices, edges of color one are added, re-routing of the loops
through the vertices occurs, and one can compute the corresponding 
pre-factors.

Also the evaluation presented in the paper can be applied to the 
q-deformed spin networks \cite{MaSmo}, \cite{BoMaSmo}. In the q-deformed 
theory all the formulae we used have their counterparts.
Thus in that case one should simply replace the formulae we derived with
the corresponding q-deformed versions \cite{KaLi}.

As we discussed before, to 
get rid of the background dependence of the pre-factor we have to know 
the angle between each pair of tangents, which requires modification 
of the regularization. Another set of problems occurs when we try to make 
sense of the square root involved in the definition of the Hamiltonian. 
There are certain proposals in this direction: One \cite{BoRoSmo} 
is to use an expansion in inverse powers of $\Lambda$ and to find 
an approximation to the square root. Another proposal \cite{Ro1} 
involves different procedure, which also allows the square root to be 
expressed as a series. In 
any case we have to settle this set of problems before trying to solve 
for the eigenvalues of the Hamiltonian operator. 

\vspace{.5 in}

This work would be impossible without the numerous discussions I had  
with Lee Smolin and Carlo Rovelli. I also would like also to thank Don 
Neville and Seth Major for their comments and criticism and the Center 
for Gravitational Physics and Geometry where this work was started, for 
the hospitality.

\newpage
 
\appendix
\section*{Basic formulae of the recoupling theory.}
One of the main results of the recoupling theory of colored knots and 
links with trivalent vertices \cite{KaLi} is the computation of the 
Kauffman bracket for
different framed spin networks. The framing refers to the fact that in the 
computations one keeps track of over- and under-crossing. In our work we do not
make this distinction so we use a simplified version of the recoupling theory, 
namely we replace the deformation parameter $q$ by its ``classical" value $-1$, 
relevant to our case. As a result in most of the formulae from 
\cite{KaLi} the q-deformed (or quantum) integers are replaced by ordinary 
ones. We list here the basic formulae, 
which we use, with the mentioned corrections made. For more details see 
\cite{KaLi}.

\begin{itemize}
\item The basic relation in this theory expresses the relation between the
different ways in which three angular momenta, say $j_{1}$, $j_{2}$,
and $j_{3}$ can couple to form a fourth one, $j_4$. The two
possible recouplings are related by the formula:

\begin{equation}
\kd{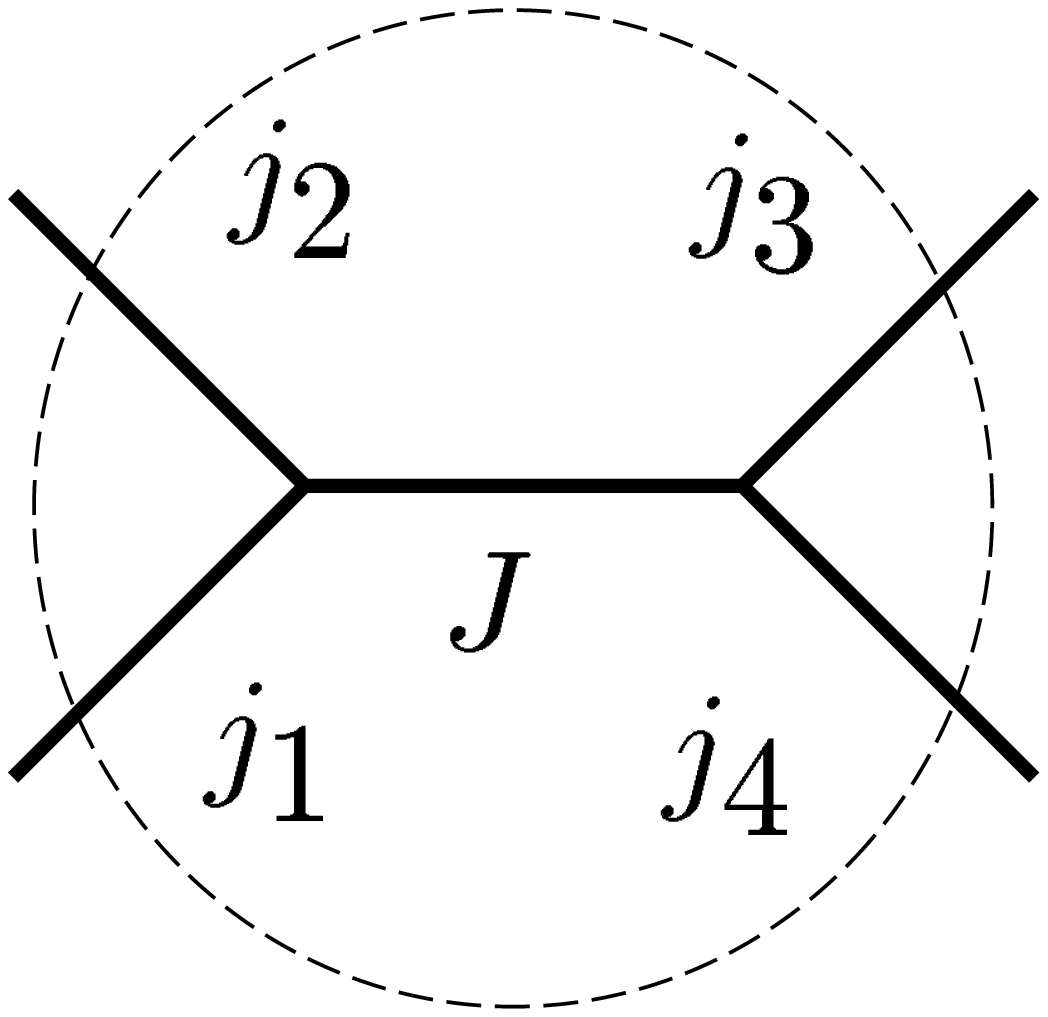} = \sum_I
\left\{ \begin{array}{ccc} j_1 & j_2 & I \\ j_3 & j_4 &J
\end{array} \right\} \kd{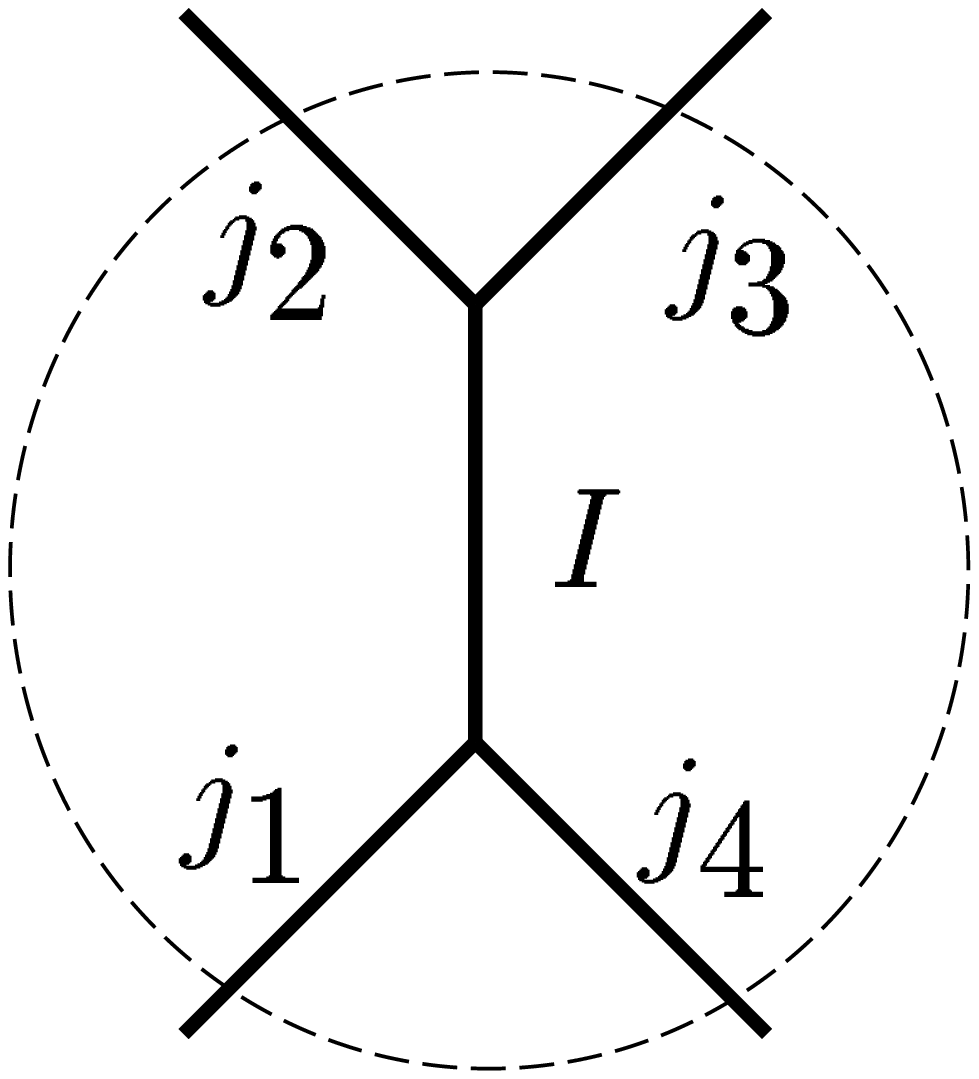}
\label{recop}
\end{equation}
where on the right hand side is the $q$-6j symbol, as defined in
\cite{KaLi} for the value of the deformation parameter $q$ equal to -1.
Again the dashed line denotes the fact that the recoupling
occurs in a region which shrinks to a point; it is not extended in space.

\item Closed loops which have been shrunk to a point are replaced by 
their loop value, which is (for a single loop with zero-self-linking) equal to 
$-2$.  The evaluation of a single unknotted loop with color $n$ is
\cite{KaLi}:

\begin{equation}
\pic{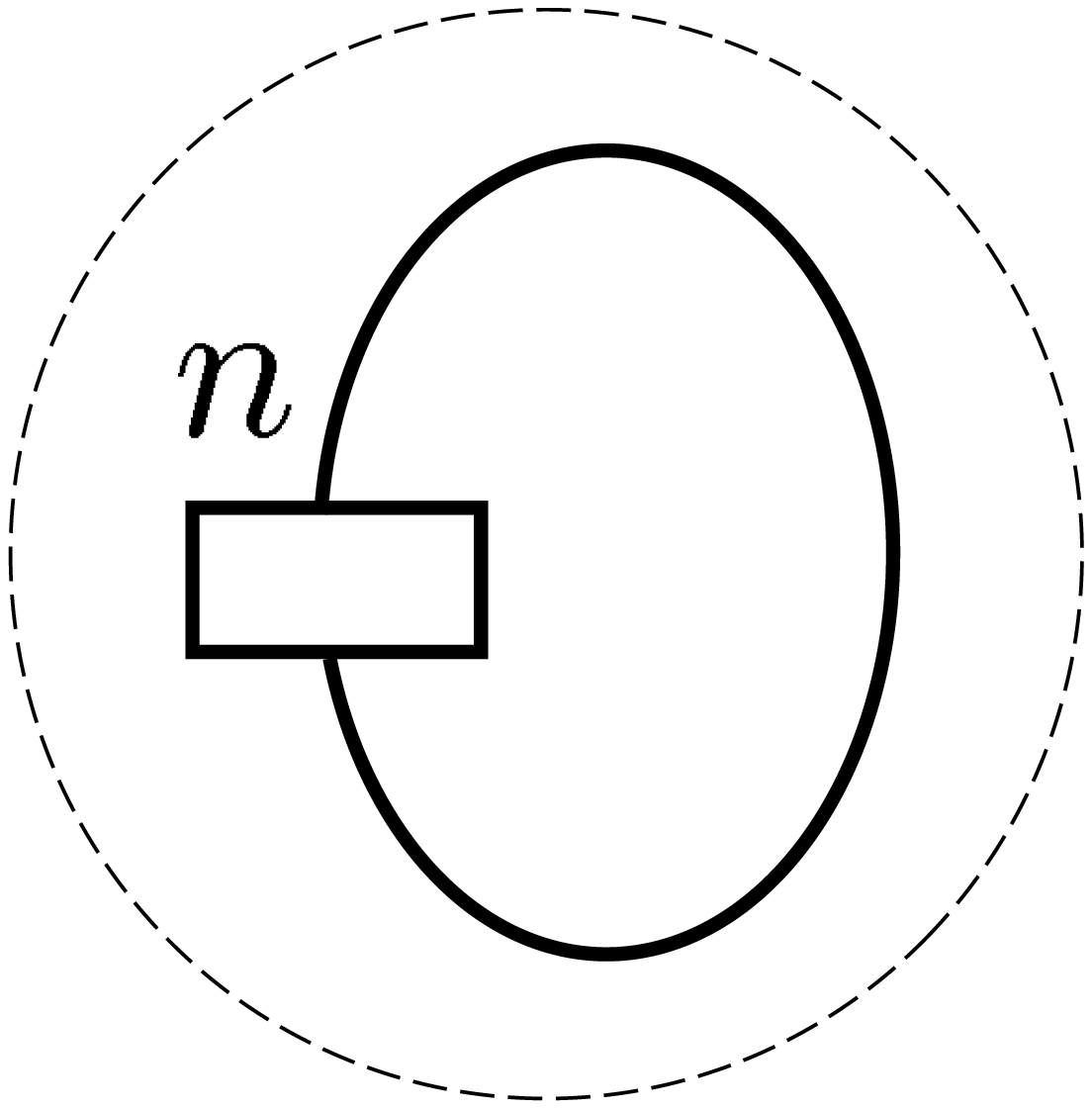}{40}{-15}{0}{0}  = \Delta_{n} = (-1)^n (n+1).
\end{equation}
The small rectangle in the above diagram denotes the antisymmetrization 
of the $n$-line.

\item The next graph we consider is the ``bubble" diagram. Upon shrinking the 
``bubble'', this diagram will reduce to a single edge so 
the evaluation will be different from zero only if the colors of both 
ends of the ``bubble'' are the same. Thus the ``bubble'' diagram equals 
some numerical factor 
times a single edge. By closing the free ends of the diagram it is 
straightforward to show that:

\begin{equation}
\kd{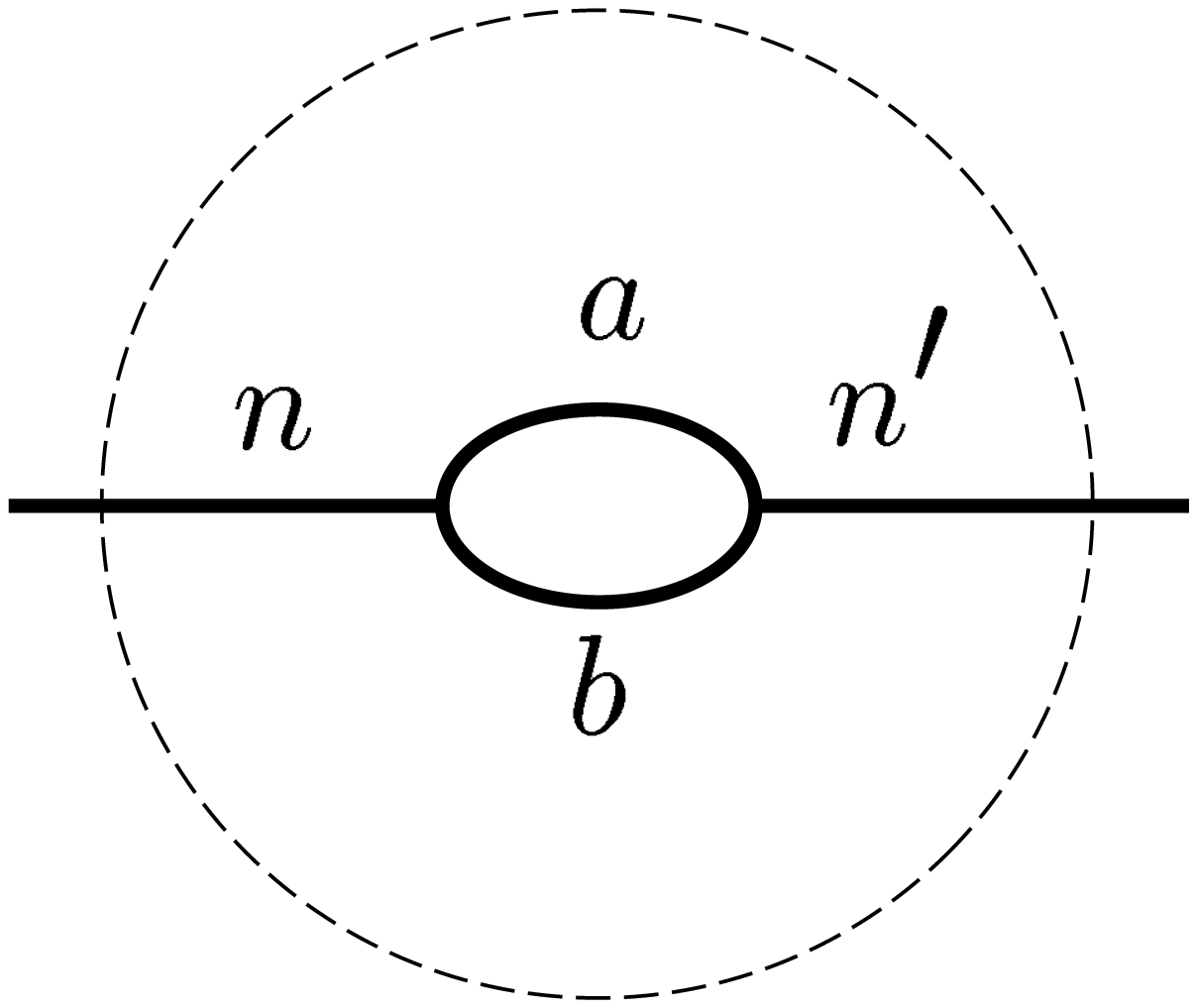} = \delta_{nn'}{ (-1)^{n} \theta(a, b, n) \over n+1  }
\kd{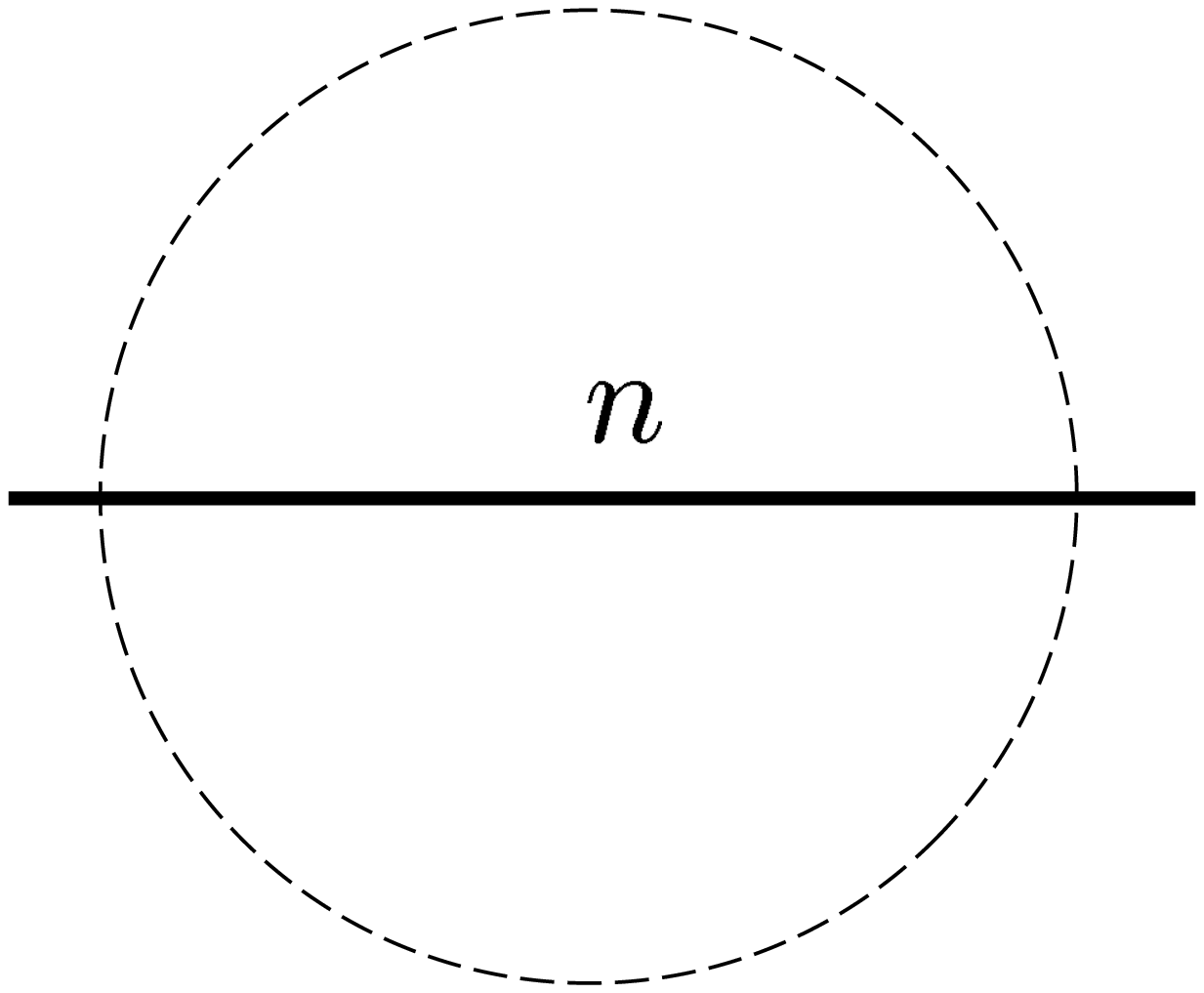}
\end{equation}
in which the function $\theta(a, b, n)$ is given, in general, by
\f\label{theta_net}
\theta(m,n,l)= \kd{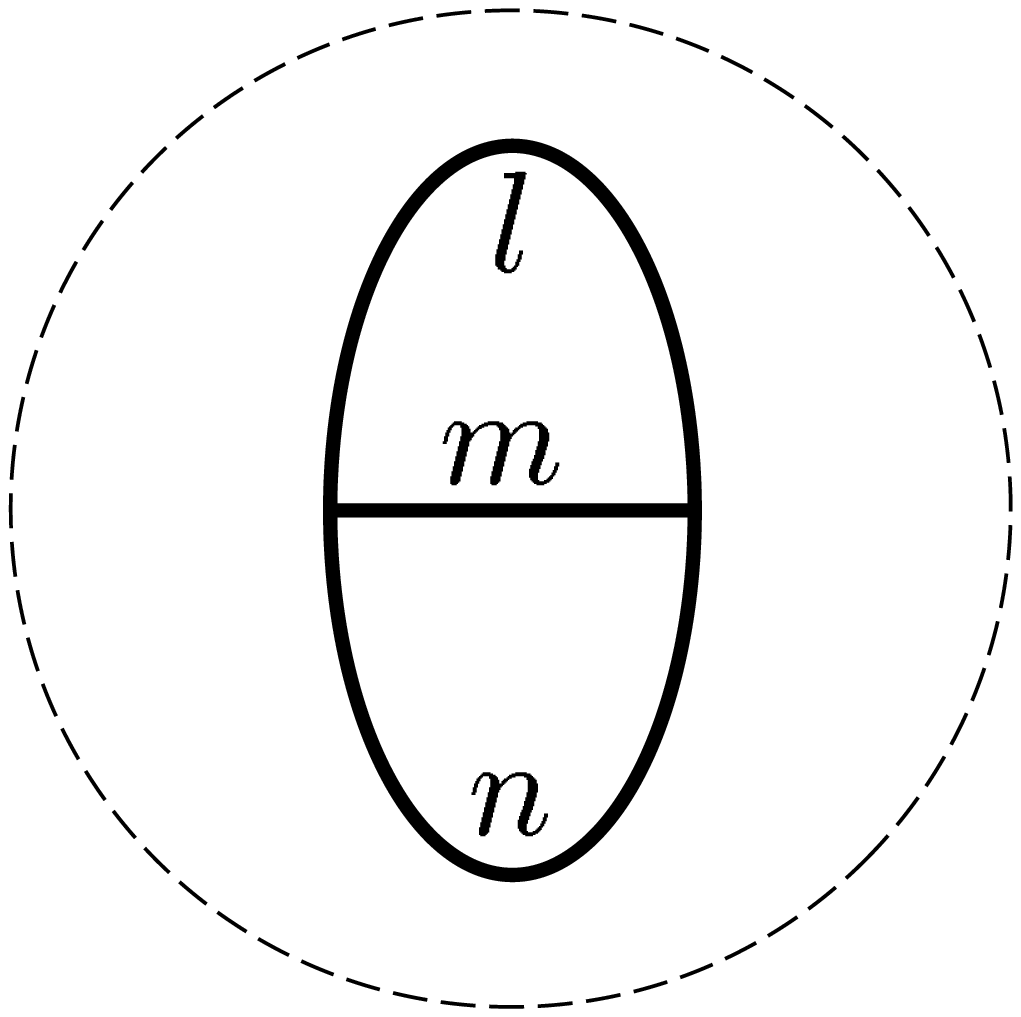} =
(-1)^{(a+b+c)}{(a+b+c+1)!a!b!c! \over (a+b)!(b+c)!
(a+c)!}
\ff
where $a+b=m$, $a+c=n$, and $b+c=l$.

\item A basic element in most of the formulae of the recoupling theory 
of colored graphs 
is the 6j symbol.  It is defined through the so called Tetrahedral net 
via the following relation

\begin{equation}\label{6Tet}
\left\{ \begin{array}{ccc} a &b & e\\ c & d & f
\end{array} \right\} = {(-1)^{e} (e+1 ){\rm Tet} \left[ 
\begin{array}{ccc} a &b & e\\ c & d & f \end{array} \right] 
 \over \vartheta ( a, d, e) \vartheta ( c, b, e) }.
\end{equation}

The Tetrahedral net is represented  by the following diagram

\[
\kd{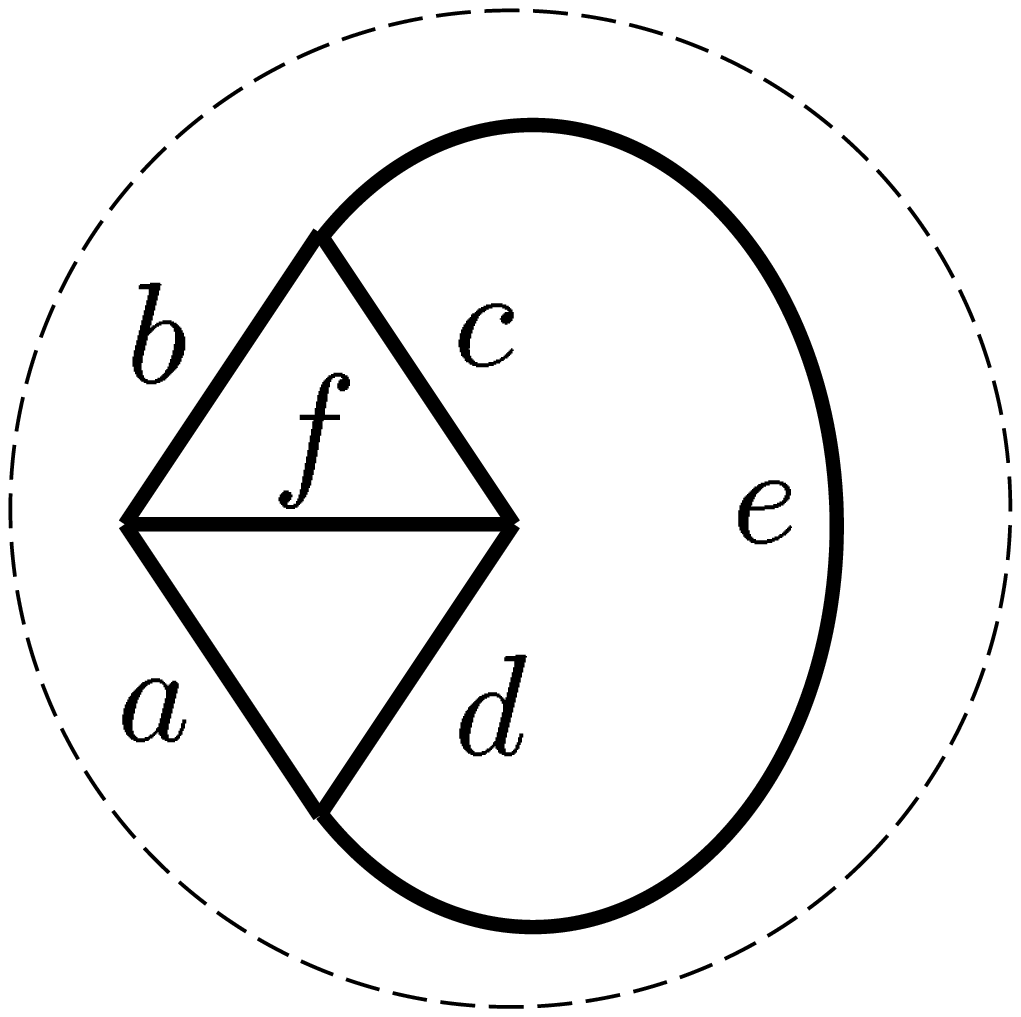} = 
{\rm Tet} \left[ \begin{array}{ccc} a &b & e\\ c & d & f
\end{array} \right] .
\]
Upon evaluation the Tetrahedral net yields:
\[
{\rm Tet} \left[ \begin{array}{ccc} a &b & e\\ c & d & f
\end{array} \right]
= \frac{{\cal I}}{{\cal E}} \sum_{m\leq s \leq M}
 \frac{ (-1)^{s} (s+1)!}{\prod_i  ~(s-a_i)!~\prod_j ~(b_j-s)! } ~~,
\]
where
\[
\begin{array}{rclcrcl} 
   a_1&=&{1  \over 2}{(a+d+e)},
   &\qquad&b_1&=&{1  \over 2}{(b+d+e+f)}, 
\end{array}
\]
\[
\begin{array}{rclcrcl}
   a_2&=&{1  \over 2}{(b+c+e)} ,
   &\qquad&b_2&=&{1  \over 2}{(a+c+e+f)}, 
\end{array}
\]
\[
\begin{array}{rclcrcl}
   a_3&=&{1  \over 2}{(a+b+f)} ,
   &\qquad&b_3&=&{1  \over 2}{(a+b+c+d)}, 
\end{array}
\]
\[
\begin{array}{rcl}
   a_4&=&{1  \over 2}{(c+d+f)} ,      
\end{array}
\]

\[
\begin{array}{rclcrcl} 
m &=&{\rm max}\{ a_i \},   & &   M & = &{\rm min}\{ b_j \}, \\[2 mm]
{\cal E} &=& a! ~b! ~c! ~d! ~e! ~f!,
& &{\cal I} &=& \prod_{ij} (b_j-a_i)! ~. 
\end{array}
\]

These formulae are sufficient for the computations performed in
the paper. In \cite{KaLi} one can find a detailed derivation of all of them.

\end{itemize}

\end{document}